\documentclass{aa}  

\usepackage{graphicx}
\usepackage{subcaption}
\usepackage{txfonts}
\usepackage{float}
\usepackage{comment}
\usepackage{gensymb}
\usepackage[debug,colorlinks=true,linkcolor=blue,citecolor=blue,urlcolor=blue]{hyperref}
\usepackage{placeins}

\newcommand{\doublehline}{\hline\hline\\[-1.5ex]}

\begin{document}

   \title{Semi-Supervised Rotation Measure Deconvolution and its application to MeerKAT observations of galaxy clusters}
   
\titlerunning{Semi-Supervised Rotation Measure Deconvolution}

   \subtitle{}

   \author{V. Gustafsson
          \inst{1}
          \and
          M. Br\"uggen \inst{1}
          \and
          T. En{\ss}lin\inst{2}
          }

   \institute{$^1$ Hamburger Sternwarte, Universität Hamburg, Gojenbergsweg 112, 21029 Hamburg, Germany\\
   $^2$ Max-Planck Institut f\"ur Astrophysik, Karl-Schwarzschild-Str. 1, 85748 Garching, Germany}

   \date{Received \today}

 
\abstract
{Faraday rotation contains information about the magnetic field structure along the line of sight and is an important instrument in the study of cosmic magnetism. Traditional Faraday spectrum deconvolution methods such as RMCLEAN face challenges in resolving complex Faraday dispersion functions and handling large datasets.}
{We develop a deep learning deconvolution model to enhance the accuracy and efficiency of extracting Faraday rotation measures from radio astronomical data, specifically targeting data from the MeerKAT Galaxy Cluster Legacy Survey (MGCLS).}
{We use semi-supervised learning, where the model simultaneously recreates the data and minimizes the difference between the output and the true signal of synthetic data. Performance comparisons with RMCLEAN were conducted on simulated as well as real data for the galaxy cluster Abell 3376.}
{Our semi-supervised model is able to recover the Faraday dispersion with great accuracy, particularly for complex or high-RM signals, maintaining sensitivity across a broad RM range. The computational efficiency of this method is significantly improved over traditional methods. Applied to observations of Abell 3376, we find detailed magnetic field structures in the radio relics, and several AGN. We also apply our model to MeerKAT data of Abell 85, Abell 168, Abell 194, Abell 3186 and Abell 3667.}
{We have demonstrated the potential of deep learning for improving RM synthesis deconvolution, providing accurate reconstructions at high computational efficiency. In addition to validating our data against existing polarization maps, we find new and refined features in diffuse sources imaged with MeerKAT.}


    \maketitle
%
\section{Introduction}

The utilization of Faraday rotation in astrophysics  spans a diverse array of cosmic phenomena, ranging from the interstellar medium of our own Galaxy to extragalactic sources. By studying the amount of Faraday rotation that polarized radio emission suffers from the source to the observer, we learn about the magnetic properties of the intervening medium. The so-called Faraday depth contains information on the strength of the magnetic field along the line of sight. The dispersion of the Faraday depth gives us information about a possible turbulent component of the magnetic field. Modern radio interferometers, such as MeerKAT \citep{Jonas_2016}, the LOw Frequency ARray (LOFAR; \cite{van_Haarlem_2013}) and the Australian Square Kilometre Array Pathfinder (ASKAP; \cite{Johnston_2007}), allow us to study the radio sky across a wide frequency band, while achieving unprecedented precision in polarization measurements. 

The MeerKAT radio interferometer \citep{Jonas_2016}, situated in South Africa, is particularly suited to polarization studies, owing to its large bandwidth. The dense inner configuration of the MeerKAT array provides high sensitivity to extended emission, while its longest baseline, spanning 7698 m, offers high angular resolution. At the central frequency of 1283 MHz, the largest resolved angular size is about 27.5$\arcmin$ together with a nominal resolution of about 6$\arcsec$.

Multiple methods of decomposing the Faraday rotating signal have been proposed over the years, e.g. Faraday Rotation Measure (RM) synthesis (\cite{Burn_1966}, \cite{Brentjens_2005}, \cite{Bell_2012}), wavelet decomposition \citep{Frick_2010}, compressive sampling (\cite{Li_2011}, \cite{Andrecut_2012}) and $QU$-fitting (\cite{Farnsworth_2011}, \cite{O'Sullivan_2012}, \cite{Ideguchi_2014}). For a comparison between the algorithms, see \cite{Sun_2015}. Parametric fits for the spectra of Faraday depth and Stokes $Q$ and $U$ typically assume a Gaussian random distribution for the components of the turbulent magnetic field. Model-free descriptions have been developed, e.g. by \cite{Van_Eck_2018}.

In this work we will focus on RM synthesis which decomposes the polarized emission into its constituent Faraday rotating components in a computationally cost-effective way. However, due to the limited coverage in frequency space, uncertainties and false positives in the form of side-lobes arise in the Fourier space. Methods such as RMCLEAN \citep{Heals_2009} reconstruct the RM synthesis signal iteratively, gradually adding point sources to the spectra until the residual is below a specified threshold. While this point source assumption is valid in the case where the emitted synchrotron radiation is separated from the Faraday rotating medium, it breaks down for regions where the radiation is emitted in a magnetized plasma. Examples of this include Galactic diffuse emission, supernova remnants, nearby galaxies and diffuse radio emission in galaxy clusters \citep{Sun_2015}.

 Prior work has investigated the application of deep learning techniques to classify the complexity of RM  spectra (e.g., \cite{Brown_2019}, \cite{Alger_2021}), however, deep learning has not been applied to the deconvolution of RM spectra. In analogy to radio interferometry, where the issue originates from an incomplete UV-space, recent advancements in deep learning have addressed similar issues by methods such as UV-space completion \citep{Schmidt_2022} and direct image reconstruction using diffusion models \citep{Wang_2023}.

In this study, we propose a novel approach for the deconvolution of RM spectra employing deep learning, by training on data from the The MeerKAT Galaxy Cluster Legacy Survey \citep{Knowles_2022}, together with simulated data, generated to match the distribution of the observational data.

This paper is organized as follows: In Sec.~\ref{sec:RM-synthesis} we describe the basics of RM synthesis. In Sec.~\ref{sec:observation} we describe the observations used in this work. Section \ref{sec:deep-learning} explains the the deep learning model, including the data pre-processing, the neural network architecture and training. In Sec.~\ref{sec:results} we present the results from simulated and observational data and compare our results to RMCLEAN. The discussion and conclusions are given
in Sec.~\ref{sec:discussion} and ~\ref{sec:summary}.

\section{Faraday Rotation Measure Synthesis}
\label{sec:RM-synthesis}

In analogy to radio interferometry, the Faraday dispersion relation can be expressed by the measurement equation, relating the Faraday depth $\phi$, to the spectral dimension $\lambda^2$. Following the notations from \cite{Burn_1966} for the Faraday dispersion function $F(\phi)$ and the complex polarized intensity $P(\lambda^2)$, the most simple form of the measurement equation reads

\begin{equation}
    \label{eq:measurement_eqation}
    P(\lambda^2) = \int_{-\infty}^{\infty} F(\phi)e^{2i\lambda^2\phi}d\phi + n(\lambda^2) ,
\end{equation}
where $n(\lambda^2)$ is noise, which for this work is assumed to be uncorrelated in $\lambda^2$ and Gaussian distributed in the real and imaginary parts. Additional terms, such as channel weights and a channel dependent sensitivity window \citep{Pratley_2020} that has been proposed to account for channel-averaging effects, are not used in this work. As we are working with a limited bandwidth, in both $\lambda^2$ and $\phi$, Eq.~\ref{eq:measurement_eqation}, reduces to the discrete inverse Fourier transform

\begin{equation}
    \label{eq:DFT}
    \Tilde{P}(\lambda^2) = \sum_i^N F(\phi_i)e^{2i\lambda^2\phi_i} + n(\lambda^2) ,
\end{equation}
or in matrix form

\begin{equation}
    \label{eq:matrix_equation}
    y = \Phi x + n ,
\end{equation}
where $\Phi$ is the measurement operator of RM synthesis. Given the increasingly large data sizes generated by modern interferometers, data is often averaged over frequency. While this process increases the signal-to-noise ratio and reduces the computational cost of both calibration and imaging, it also carries the risk of bandwidth depolarization. For Faraday rotating signals, where the real and imaginary parts are a series of sinusoidal waves, the averaging process can smear out the phase information across frequency channels, leading to a loss of polarization information. This effect becomes more pronounced for high rotation measures or low frequencies, where the signal rotates rapidly. This can average the resulting signals to near zero. As a result, the maximum Faraday depth to which one has more than 50\% sensitivity is approximately $\|\phi_{\rm max}\|=\sqrt{3} / \delta\lambda^2$.

Here we will write the channel-averaging operator as

\begin{equation}
    \label{eq:averaging}
    \mathcal{A}y_i = \frac{1}{N}\sum_j^Ny_j ,
\end{equation}
where $y_i$ is the average signal intensity of the $i$-th channel bin, and $N$ is the number of channels in each bin.

By expressing these operations in matrix form, we can leverage the broadcasting feature of matrix multiplication, enabling efficient element-wise operations across arrays of different shapes without explicit looping. This, coupled with the parallel computing capabilities of GPUs, accelerates the execution of the algorithm for processing large-scale datasets significantly.

\section{Observation}
\label{sec:observation}

The data used in this paper has been taken from the MeerKAT Galaxy Cluster Legacy Survey (MGCLS) DR1. For imaging and calibration details see \cite{Knowles_2022}. Out of the total number of 115 surveyed galaxy clusters, 44 were imaged in full Stokes. The data products include image cubes consisting of 12 frequency channels (908-1656 MHz), within the L-band (856-1712 MHz), at the full 8$\arcsec$ and 15$\arcsec$ resolution to help recover low-surface-brightness features, such as radio relics. Table \ref{tab:spec_config} provides the spectral and RM configuration of this dataset.

\begin{table}[]
    \centering
    \caption{Spectral and RM synthesis configuration.}
    \begin{tabular}{@{}p{0.6\columnwidth}l@{}}
    \doublehline
       \hspace{5pt} Number of channels &  \hspace{5pt}\\
       \hspace{5pt} -- initial & 4096 \hspace{5pt}\\
       \hspace{5pt} -- final & 12 \hspace{5pt}\\
       \hspace{5pt} Frequency resolution $\delta\nu$ &  \hspace{5pt}\\
       \hspace{5pt} -- initial & 209 kHz \hspace{5pt}\\
       \hspace{5pt} -- final & 62 MHz (mean) \hspace{5pt}\\
       \hspace{5pt} Wavelength squared bandwidth $\Delta\lambda^2$ & 0.076 m$^{2}$ \hspace{5pt}\\
       \hspace{5pt} -- min resolution $\delta\lambda^2_{min}$ & 0.0019 m$^{2}$ \hspace{5pt}\\
       \hspace{5pt} -- max resolution $\delta\lambda^2_{max}$ & 0.0168 m$^{2}$ \hspace{5pt}\\
       \hspace{5pt} Faraday depth bandwidth $\Delta\phi$ & 512 rad m$^{-2}$ \hspace{5pt}\\
       \hspace{5pt} Faraday depth sampling $\delta\phi$ & 1 rad m$^{-2}$ \hspace{5pt}\\
       \hline
    \end{tabular}
    \label{tab:spec_config}
\end{table}

\begin{figure}[ht]
    \centering
    \includegraphics[width=\columnwidth]{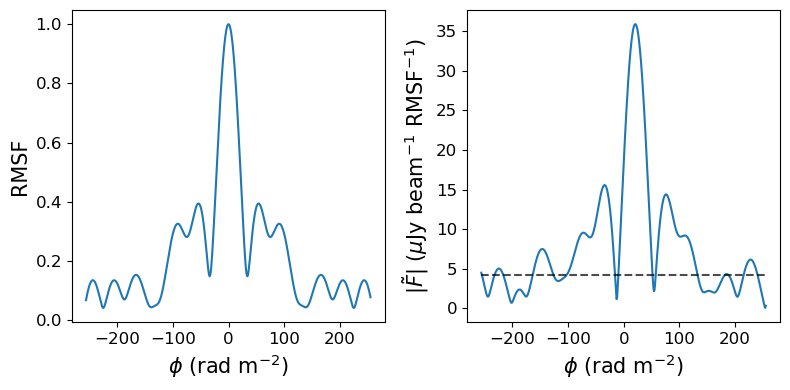}
    \caption{Magnitude of the RMSF for the MGCLS band (left) and an example Faraday spectrum taken from a pixel in the eastern relic (right). The estimated noise level is shown as a black dashed line.}
    \label{fig:RMSF}
\end{figure}

The 12 spectral channels of the dataset provide an RM sensitivity up to 173 rad m$^{-2}$, where the sensitivity is reduced to half. However, as we will see in Sec.~\ref{sec:simulations}, using deep learning, we can detect higher values of RM by learning the sensitivity window of our observation. Consequently, the selected RM search range is set to [-256, 256) rad m$^{-2}$, with a sampling in Faraday space of 1 rad m$^{-2}$. Due to the limited bandwidth of our observation, each point in $\phi$ will be convolved with the Rotation Measure Spread Function (RMSF), shown in Figure \ref{fig:RMSF}, together with an example Faraday spectrum from the Abell 3376 dataset.

\begin{figure*}
    \centering
    \includegraphics[width=\textwidth]{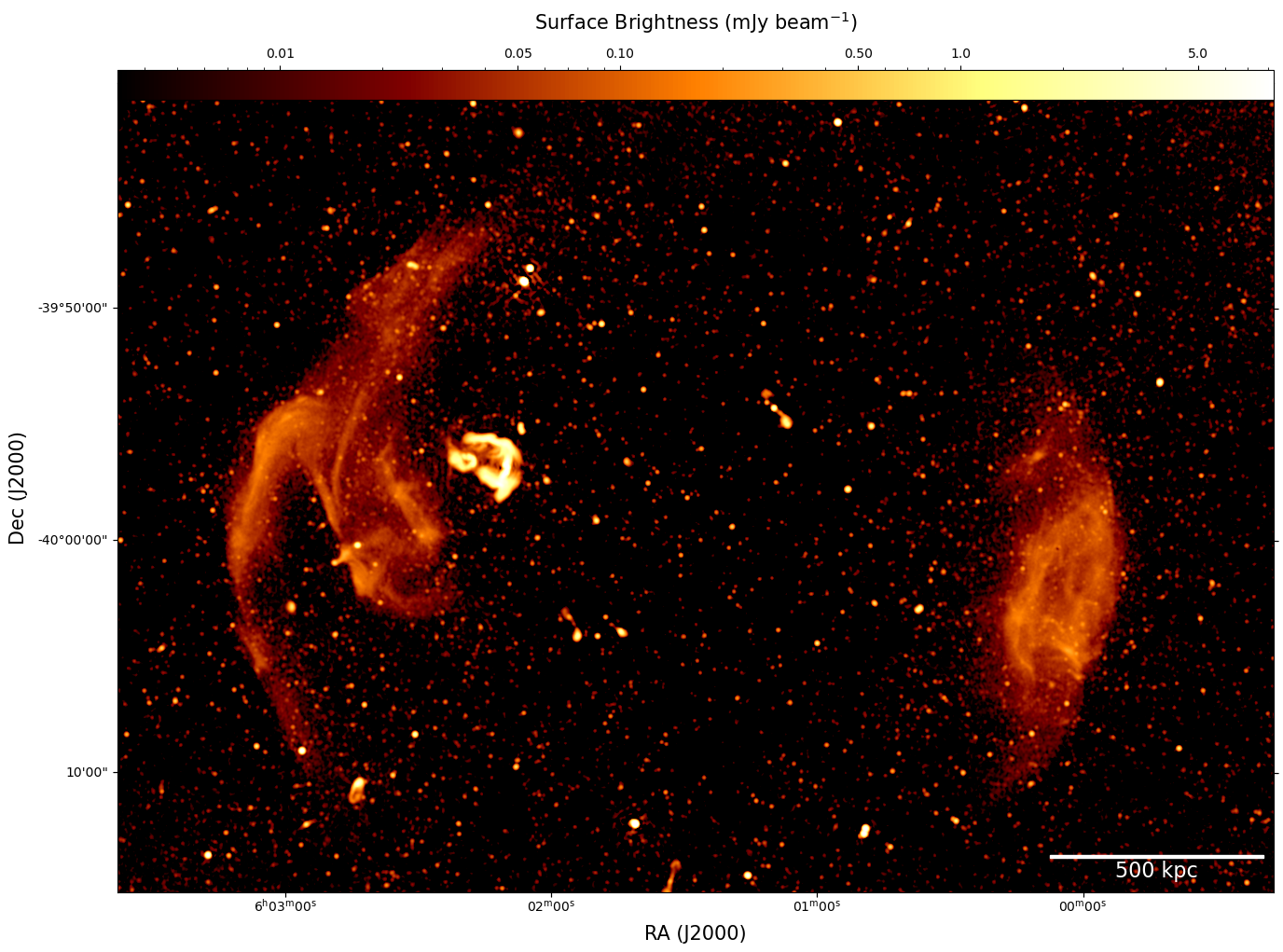}
    \caption{Total intensity map (908-1656 MHz) of Abell 3376 at a resolution of $8\arcsec\times8\arcsec$. The image has a noise level of 13 $\mu$Jy beam$^{-1}$.}
    \label{fig:Stokes_I}
\end{figure*}

The main focus in this paper is the Abell 3376 cluster, which has previously been studied by, e.g., \cite{Kale_2012}, \cite{George_2015}, \cite{Chibueze_2023}. The studies mostly focused on the radio relics. In \cite{Hu_2024}, synchrotron intensity gradient (SIG) mapping was used to infer the magnetic field orientation in the radio relics. However, no studies have yet been conducted with MeerKAT in full polarization. Fig.~\ref{fig:Stokes_I} show the total intensity map at an angular resolution of 8$\arcsec$. The radio sources in Abell 3376 have a large angular size which means that the observations reveal a great level of detail.

This cluster harbors two radio relics that extend for Megaparsecs. No radio halo has been detected in this cluster. The orientation of the cluster relics, together with the extended X-ray emission stretching in the northwest-southeast direction \citep{Kale_2012}, suggests that Abell 3376 is a merging cluster. 

Furthermore, the cluster contains several radio galaxies and active galactic nuclei (AGN). Most notably close to the eastern relic, with jets bent by $\sim90$ degrees from their original direction, which is suggested to be caused by the cluster magnetic field \citep{Chibuezu_2021}.

In appendix \ref{sec:clusters} we briefly go over the results of applying the deep learning model to five more datasets from the MGCLS DR1, and compare them with previous studies.

\section{Deep Learning Deconvolution}
\label{sec:deep-learning}

The neural network model aims to invert Eq.~\ref{eq:matrix_equation}, which generally lacks invertibility. However, we can approximate a highly nonlinear function tailored to this specific task by fine-tuning the parameters of a neural network. The parameter optimization process uses the ADAM optimizer \citep{Kingma_2017}, aiming to minimize the mean squared error (MSE) loss between the observed data and the result obtained after applying the measurement and channel-averaging operators to the network's predictions, as given by:

\begin{equation}
\label{eq:loss_MSE}
L = \frac{1}{N}\sum_i^N\|{y_i - \mathcal{A}\Phi \hat{x}_i}\|^2 ,
\end{equation}
where $N$ is the number of samples in each batch, $\hat{x}_i$ is the model output in $\phi$ and $y_i$ is the data in $\lambda^2$. The motivation behind using the MSE loss lies in the assumption of Gaussian-distributed noise in both the real and imaginary parts of the complex polarized intensity. This choice is based on the observation that the log likelihood function for a Gaussian distribution is directly linked to the squared Euclidean norm, as seen in the MSE formulation. This approach ensures that our loss function captures the statistical characteristics of the data while minimizing the error between predicted and observed values.

Minimizing Eq.~\ref{eq:loss_MSE} allows us to identify a solution that aligns with the observed data. However, given the problem's infinite solution space, we require a prior to guide the optimization towards physically realistic solution. Previous work, such as non-parametric $QU$-fitting \cite{Pratley_2021}, incorporate an $\ell_1$ regularization term to constrain the number of RM components, essentially adopting a CLEAN prior approach. Instead, this work implicitly includes a Faraday-thick prior by utilizing semi-supervised learning, including samples with simulated Gaussian sources in $\phi$, in the training dataset, and allowing the model access to the true signal of those samples. The model should thus find a general solution, that reconstructs the simulated sources, while also fitting the observed data. The full semi-supervised loss thus reads

\begin{equation}
    \label{eq:loss_full}
    L = \frac{\beta}{N_{\rm obs}}\sum^{N_{\rm obs}}_i\|{y_i - \mathcal{A}\Phi \hat{x}_i}\|^2 + \frac{1}{N_{\rm sim}}\sum^{N_{\rm sim}}_j\|{x_j - \hat{x}_j}\|^2 ,
\end{equation}
where the loss is divided into contributions from real samples ($N_{\rm obs}$) and simulated samples ($N_{\rm sim}$). The first term of the loss function thus compares the spectra in $\lambda^2$, while the second part compares the output $\hat{x}_j$ with the true signal $x_j$ in $\phi$. The factor $\beta$ is used for weighting the different loss terms, assuring that information is passed from the supervised learning onto the observational data. 

\subsection{Pre-Processing}
\label{pre-processing}

\begin{figure*}
    \centering
    \includegraphics[width=\textwidth]{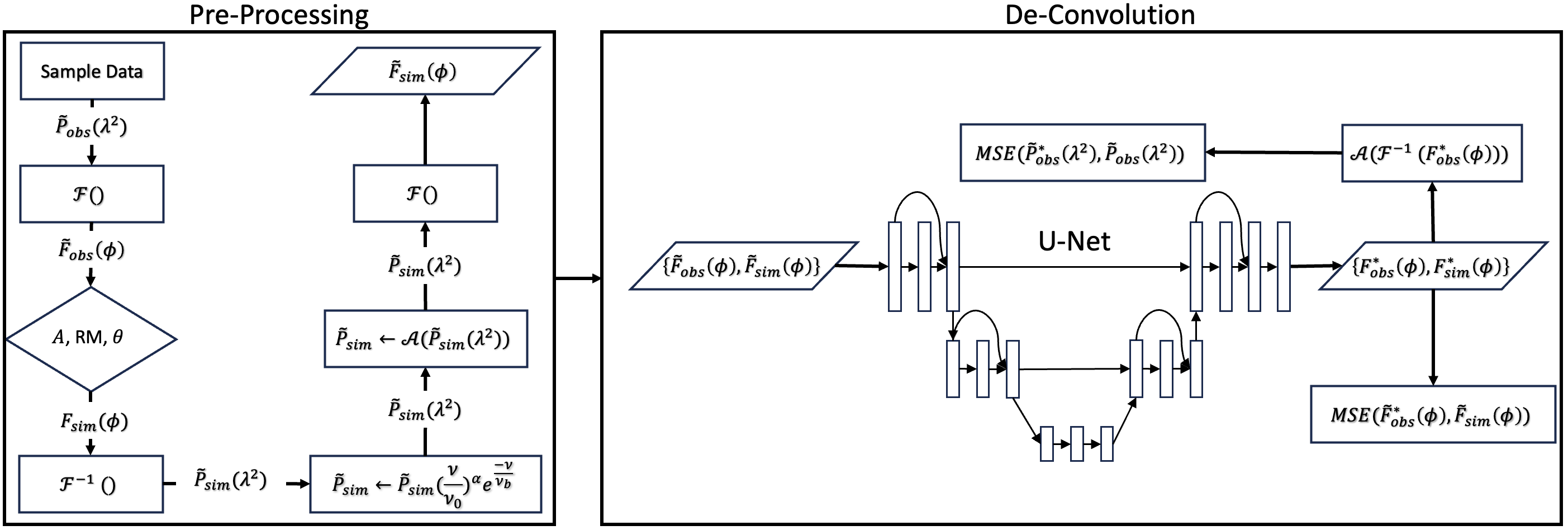}
    \caption{Flowchart illustrating our deep learning deconvolution model. The notations $\mathcal{F}$ and $\mathcal{F}^{-1}$ signify the transformations from $\lambda^2$ to $\phi$ and back, respectively. Initially, samples are extracted from the observed data and used to generate a simulated dataset with a similar distribution. Next, the two datasets are combined and fed into the neural network. After the deconvolution process, the predictions $F^*_{obs}$ and $F^*_{sim}$ are separated once more and evaluated against their corresponding targets using the MSE loss.}
    \label{fig:Model}
\end{figure*}

Creating a balanced dataset of synthetic data is a relatively simple process. However, observational data is typically unbalanced and may include outliers, which can negatively affect the training process. Among the samples, bright point sources have the most significant impact. Additionally, to avoid training the model on data predominantly comprised of noise, noise dominated samples were also excluded from the training dataset. As the polarized intensity map was significantly contaminated by foreground emission, a threshold was instead set by the noise level in total intensity $\sigma_{\text{I}}$, calculated as

\begin{equation}
    \sigma_I^2 = \frac{1}{N} \sum_i^N \sigma_{I,i}^2 ,
\end{equation}
where \( N \) is the number of channels in the Stokes $I$ cube, and \(\sigma_{I,i}\) is the noise in the \(i\)-th channel, measured from an emission-free region. Both noisy samples and bright sources were excluded by clipping the dataset to within the range of \( (3, 30)\sigma_{\text{I}} \). The full pre-processing and training steps are shown in Fig. \ref{fig:Model}.

\begin{table}[]
    \centering
    \caption{Source parameters for creating a simulated dataset that matches the data distribution. The amplitude $A$, RM, intrinsic polarization angle $\theta$ are drawn from the data $\mathcal{D}$. The number of sources, full width at half max FWHM, spectral index $\alpha$, and breaking frequency are drawn from uniform distributions.}
    \begin{tabular}{@{}p{0.7\columnwidth}l@{}}
        \doublehline
        \hspace{5pt} Number of sources & $U(1,5)$\hspace{5pt}\\
        \hspace{5pt} Amplitude $A$ ($\mu$Jy) & $\sim\mathcal{D}$\hspace{5pt}\\
        \hspace{5pt} Rotation Measure RM (rad m$^{-2}$) & $\sim\mathcal{D}$ \hspace{5pt}\\
        \hspace{5pt} Intrinsic polarization angle $\theta$ (radians) & $\sim\mathcal{D}$ \hspace{5pt}\\
        \hspace{5pt} FWHM (rad m$^{-2}$) & $U(1,10)$ \hspace{5pt}\\ 
        \hspace{5pt} Spectral index $\alpha$ & $U(-3,0)$ \hspace{5pt}\\ 
        \hspace{5pt} Breaking frequency $\nu_{\text{b}}$ (MHz) & $U(1,10)$ \hspace{5pt}\\
        \hline
    \end{tabular}
    \label{tab:sim_params}
\end{table}

After sampling the observed data, $\Tilde{P}_{\rm obs}(\lambda^2)$, we extracted the peak flux $A$, RM, and intrinsic polarization angle $\theta$ from the RM synthesis signals $\Tilde{F}_{\rm obs}(\phi)$. The full set of simulation parameters can be seen in Table \ref{tab:sim_params}. These parameters were then used to create one-dimensional simulated samples, each consisting of a Gaussian mixture model, with one to five Gaussian components in $\phi$. It is important to note that these simulations are a significant simplification of the true profile of Faraday spectra, and that realistic Faraday spectra are expected to exhibit non-Gaussian features. Examples of such features can be seen in \cite{Bell_2011}, where sharply peaked and asymmetric profiles are obtained from more advanced simulations.

The amplitude of each component was adjusted to conserve the total flux of the sample by dividing it by the number of components. Since the peak flux of the RM synthesis signal corresponds to the integral of an unresolved source, the amplitude was also adjusted by dividing it by the factor that relates the area of a Gaussian to its amplitude, given by $2\sqrt{\log(2)} /\sqrt{\pi}$ FWHM. The full width at half maximum (FWHM) of each Gaussian component was drawn from a uniform distribution between 1 and 10 rad m$^{-2}$. Each simulated sample was then transformed to $\lambda^2$, where each signal was multiplied by a combination of a power-law and an exponential:

\begin{equation}
    \Tilde{P} = \Tilde{P}_0 \left(\frac{\nu}{\nu_0}\right)^{\alpha} e^{-\nu / \nu_{\text{b}}} ,
\end{equation}
where $\alpha$ is the spectral index, $\nu_0$ is set to the lowest frequency in our band, 856 MHz, and $\nu_{\text{b}}$ is the breaking frequency. No noise is added at this stage for reasons discussed in section \ref{sec:a3376}. Samples were then averaged to the 12 frequency channels provided in the dataset. The simulated signals were then transformed to $\phi$, resulting in $\Tilde{F}_{\text{sim}}(\phi)$ and paired with their corresponding true signal $F_{\text{sim}}(\phi)$ to create input-target pairs. Similarly, each observed sample $\Tilde{F}_{\text{obs}}(\phi)$ was paired with its respective $\lambda^2$ spectrum $\Tilde{P}_{\text{obs}}(\lambda^2)$ to create input-target pairs for the observation dataset.

\subsection{Network Architecture}

The network architecture for this work is a one-dimensional U-Net \citep{Ronneberger_2015}. The network consists of five levels, each consisting of a ResNet block \citep{He_2015}. In the contracting path, max-pooling is used to progressively downsample the input by a factor of two after each ResNet block, while in the expanding path, nearest neighbor up-sampling is applied. Additionally, the model employs skip connections with attention gates \citep{Oktay_2018} to retain fine structural details. At the bottleneck, the downsampled input should contain most of the signal information in a compact latent space. From this high-dimensional representation, the model should reconstruct the signal, effectively performing the deconvolution process and recovering the true Faraday dispersion function. In Table \ref{tab:hyperparameters}, the set of hyperparameters used for deconvolving the Abell 3376 dataset are shown. The model uses a learning rate scheduler, monitoring Eq. \ref{eq:loss_full} during training and decreasing the learning rate by a factor if the loss is not reduced after a set number of epochs. This approach allows for starting with a relatively high learning rate, thus speeding up convergence, while also reducing the risk of overfitting.

\begin{table}[]
\caption{Hyperparameters for the deep learning deconvolution model.}
    \centering
    \begin{tabular}{@{}p{0.5\columnwidth}l@{}}
        \doublehline
        \hspace{5pt} Number of levels & 5 \hspace{5pt}\\
        \hspace{5pt} Minimum number of channels & 64 \hspace{5pt}\\
        \hspace{5pt} Convolution kernel size & 3 \hspace{5pt}\\
        \hspace{5pt} Dropout rate & 0.2 \hspace{5pt}\\
        \hspace{5pt} Batch size & 256 \hspace{5pt}\\
        \hspace{5pt} Initial learning rate & $10^{-5}$ \hspace{5pt}\\
        \hspace{5pt} -- min learning rate & $10^{-7}$ \hspace{5pt}\\
        \hspace{5pt} -- learning rate factor & 0.1 \hspace{5pt}\\
        \hspace{5pt} -- learning rate patience & 5 epochs \hspace{5pt}\\
        \hspace{5pt} Loss weighting factor $\beta$ & 10$^{-5}$\hspace{5pt}\\
        \hline
    \end{tabular}
    \label{tab:hyperparameters}
\end{table}

\subsection{Training}

\begin{figure}[ht]
    \centering
    \includegraphics[width=\columnwidth]{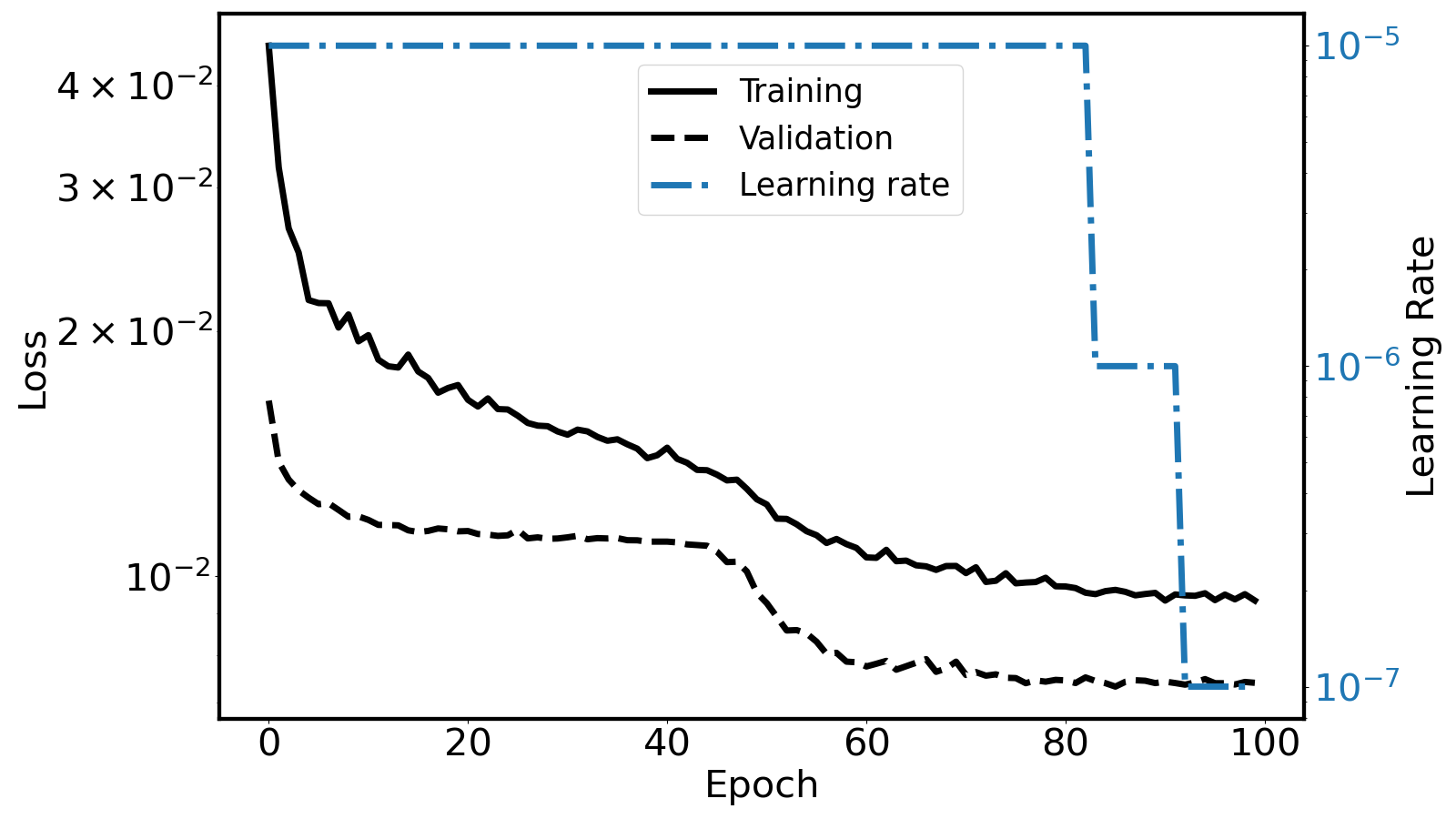}
    \caption{Loss curves and learning rate during training on the Abell 3376 dataset. The plot shows the training loss (solid line) and validation loss (dashed line) as a function of epoch, plotted on the left y-axis with a logarithmic scale. The learning rate, plotted on the right $y$-axis with a logarithmic scale, is also shown.}
    \label{fig:loss_curves}
\end{figure}

The model was trained with 80000 real and 20000 simulated samples over 100 epochs, reaching convergence after approximately 80 epochs. The dataset was divided into 80\% training, 10\% validation, and 10\% testing data. During training, the model optimizes the network parameters using the training data. The validation dataset is used to monitor the model's performance and tune hyperparameters, preventing overfitting by ensuring that the model generalizes well to unseen\footnote{Commonly used word in machine learning, data that the model has not optimized its parameters to} data. The final evaluation of the model's performance is conducted on the test dataset, which provides an unbiased assessment of its predictive accuracy. In Fig. \ref{fig:loss_curves}, Eq. \ref{eq:loss_full} is plotted as a function of epochs during the training process, together with the learning rate. We see that the model generalizes well to unseen data, as the validation loss is always lower than the training loss. This is due to the use of dropout, which randomly sets a fraction of the neurons to zero during training, in order to find a simpler and more general solution to the problem. We observe that after 40 epochs, the model encounters a local minimum, where the loss reduction slows down. However, the model manages to escape this minimum and continues to reduce the loss after approximately 10 more epochs. At around 80 epochs, the model stops reducing the loss further, prompting the learning rate scheduler to lower the learning rate.

\section{Results}
\label{sec:results}

In this section we present the results from applying our semi-supervised model first to a set of simulated data, and second to a MeerKAT observation of the galaxy cluster Abell 3376.

\subsection{Simple model of Gaussian components}
\label{sec:simulations}

In order to evaluate the recovery of the true signal, we apply our model to a data set of simulated Gaussian sources. The simulations were created similarly to section \ref{pre-processing}, but with the parameters in Table \ref{tab:sim_params2}. The noise level $\sigma_{QU}$ is added to the real and imaginary parts in $\lambda^2$, set to the square root of the number of channels, resulting in a noise level of 1 $\mu$Jy after averaging over all channels.

\begin{table}[]
    \centering
    \caption{Source parameters for a simulated testing dataset. All parameters are drawn from uniform distribution except the noise level $\sigma_{QU}$ which is the same for all samples.}
    \begin{tabular}{@{}p{0.7\columnwidth}l@{}}
        \doublehline
        \hspace{5pt} Number of sources & $ U(1,3)$\hspace{5pt}\\
        \hspace{5pt} Amplitude $A$ ($\mu$Jy) & $ U(0,50)$ \hspace{5pt}\\
        \hspace{5pt} Rotation Measure RM (rad m$^{-2}$) & $ U(-200,200)$ \hspace{5pt}\\
        \hspace{5pt} Intrinsic polarization angle $\theta$ (radians) & $ U(-\pi/2, \pi/2)$ \hspace{5pt}\\
        \hspace{5pt} FWHM (rad m$^{-2}$) & $ U(1,10)$ \hspace{5pt}\\
        \hspace{5pt} Spectral index $\alpha$ & $ U(-3,0)$ \hspace{5pt}\\ 
        \hspace{5pt} Breaking frequency (GHz) $\nu_{\text{b}}$ & $ U(1,10)$ \hspace{5pt}\\
        \hspace{5pt} $\sigma_{QU}$ ($\mu$Jy) & $\sqrt{N_{\rm channels}}$ \hspace{5pt}\\ 
        \hline
    \end{tabular}
    \label{tab:sim_params2}
\end{table}

\begin{figure*}
    \centering
    \includegraphics[width=\textwidth]{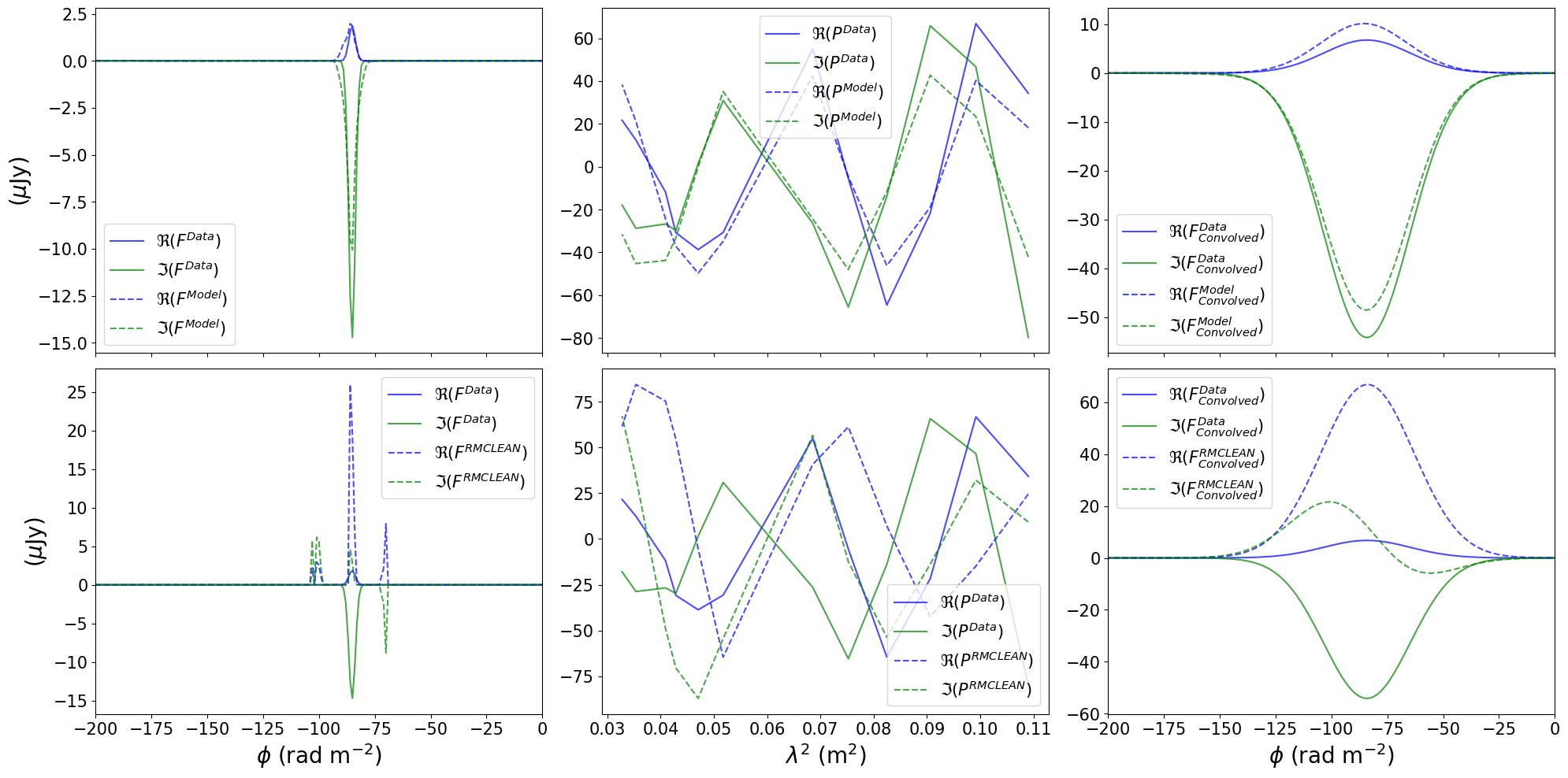}
    \caption{Comparison between deep learning model output (top) and RMCLEAN (bottom). The outputs are shown in $\phi$ (left) and $\lambda^2$ domain (middle). While the deep learning model fits the data with Gaussian shaped peaks, RMCLEAN creates a number of Faraday-thin components in $\phi$. In the right plot the outputs are tapered to the theoretical resolution of 43 rad m$^{-2}$.}
    \label{fig:RMCLEAN_1}
\end{figure*}

\begin{figure*}
    \centering
    \includegraphics[width=\textwidth]{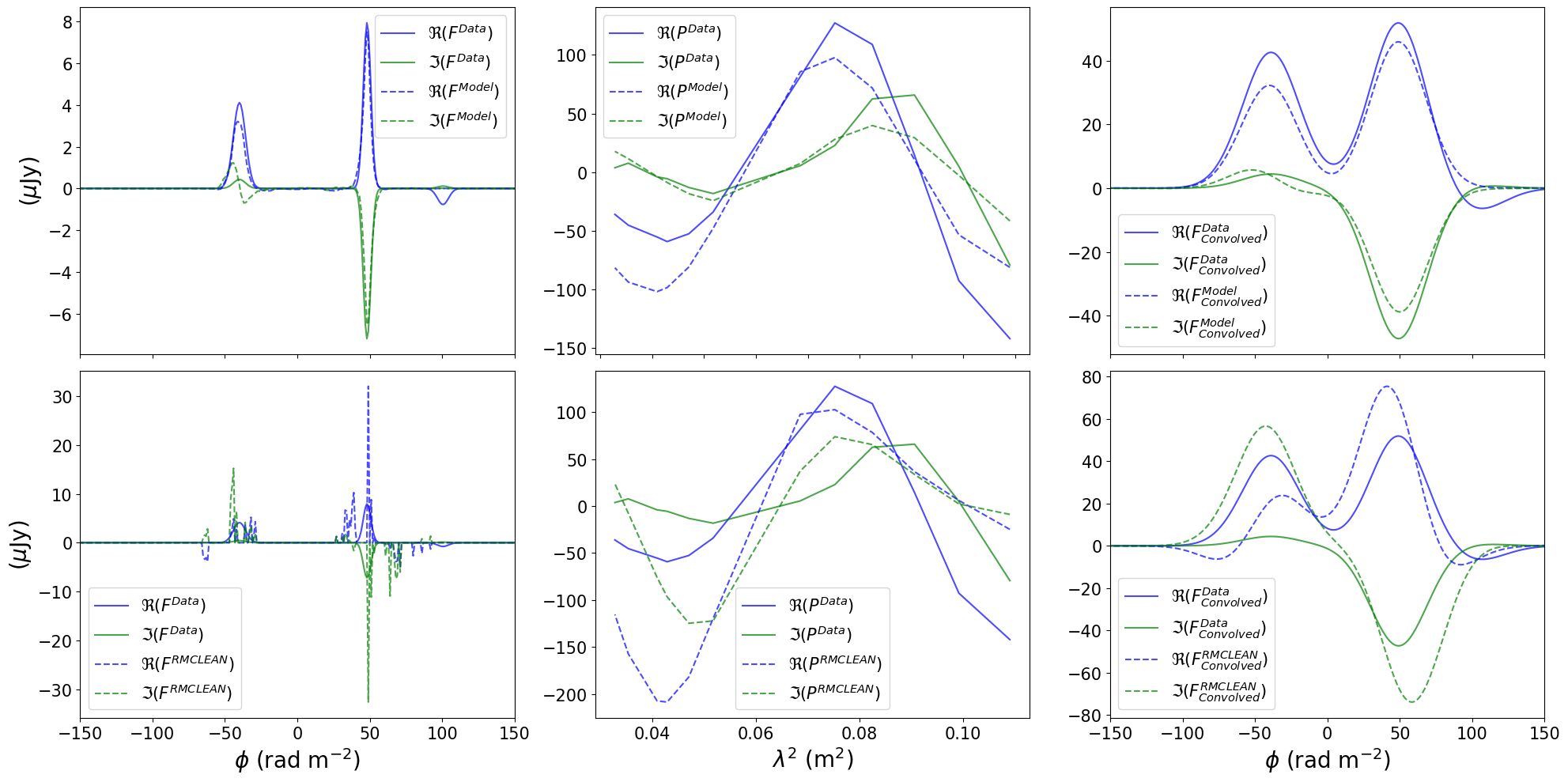}
    \caption{As Fig.~\ref{fig:RMCLEAN_1}, but with an assumed more complex Faraday spectrum.}
    \label{fig:RMCLEAN_2}
\end{figure*}

In Fig.~\ref{fig:RMCLEAN_1} and \ref{fig:RMCLEAN_2}, we compare the performance of the deep learning model and RMCLEAN. Fig. \ref{fig:RMCLEAN_1} presents a case with a single source located at $\phi = -85$ rad m$^{-2}$. Our model successfully recovers the signal and aligns well with the data in $\lambda^2$. While there are minor differences between the true source and the output, they are reduced when the resolution is set to the theoretical limit.

RMCLEAN iteratively adds point sources until a specified threshold is reached, typically just before the model starts cleaning the noise. In Fig.~\ref{fig:RMCLEAN_1}, RMCLEAN places point sources at strong emission locations. However, instead of accurately recovering the true signal, it overestimates the flux at these points, leaving regions near the peak flux blank. To reproduce the Faraday thick component, RMCLEAN also places sources further from the peak, resulting in a spread that does not correspond to the true distribution.

A more complex Faraday spectrum is shown in Fig. \ref{fig:RMCLEAN_2}, with three sources with different parameters as listed in Table \ref{tab:sim_params2}. While the deep learning model accurately recovers the strong sources, it fails to detect the weaker source which lies close to the noise level. For the complex spectrum, it becomes even more evident that the RMCLEAN assumption is not valid for extended emission in $\phi$.

A fairer comparison would be to use a multi-scale version of the CLEAN algorithm, similar to \cite{2008ISTSP...2..793C}, which adds components of gradually increasing scale sizes to the model image. As no such implementation exists for RMCLEAN, we should not compare the deep learning model output with the clean components directly, but rather look at peak flux, RM and polarization angle of the low-resolution spectra.

\begin{figure}[ht]
    \centering
    \includegraphics[width=\columnwidth]{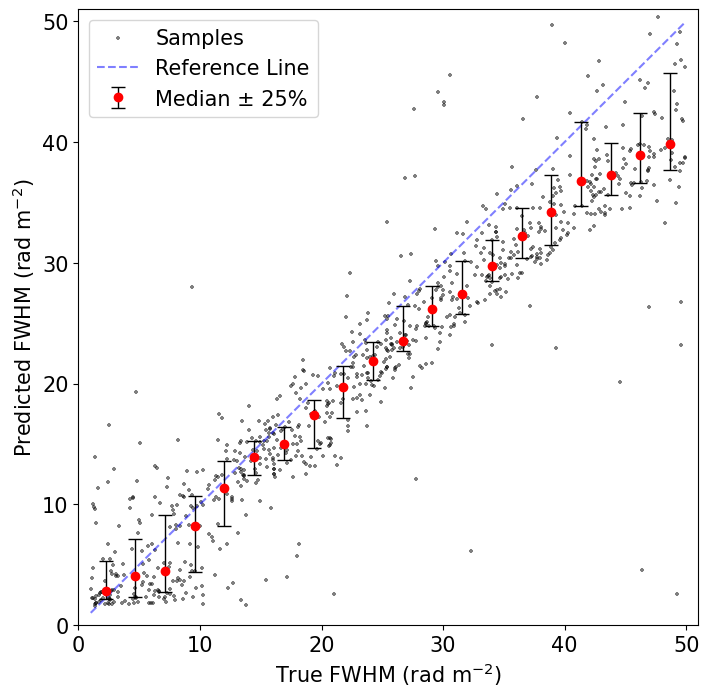}
    \caption{Predicted versus true FWHM for a test with a single Gaussian component. Predicted FWHMs are shown as black dots. The range is divided into 20 bins from which the median $\pm25\%$ are calculated and shown as red dots and errorbars respectively. A reference line where the predicted FWHM equals the true FWHM is shown as a dashed blue line.}
    \label{fig:FWHM}
\end{figure}

In order to evaluate the model's super-resolution capabilities, we conducted a test using a single Gaussian component while varying the FWHM between 1 and 50 rad m$^{-2}$. The rest of the simulation parameters are as in Table \ref{tab:sim_params2}. In Fig. \ref{fig:FWHM}, we present the comparison between the predicted FWHM and the true FWHM. We observe a clear correlation between the true and predicted FWHM, significantly below the theoretical resolution of 43 rad m$^{-2}$. However, for values below 10 rad m$^{-2}$, there is greater uncertainty, with many predictions clustering around 1 rad m$^{-2}$, which is not accurate. We also see that the FWHM is systematically underestimated throughout the entire sample.

\begin{figure}[ht]
    \centering
    \includegraphics[width=\columnwidth]{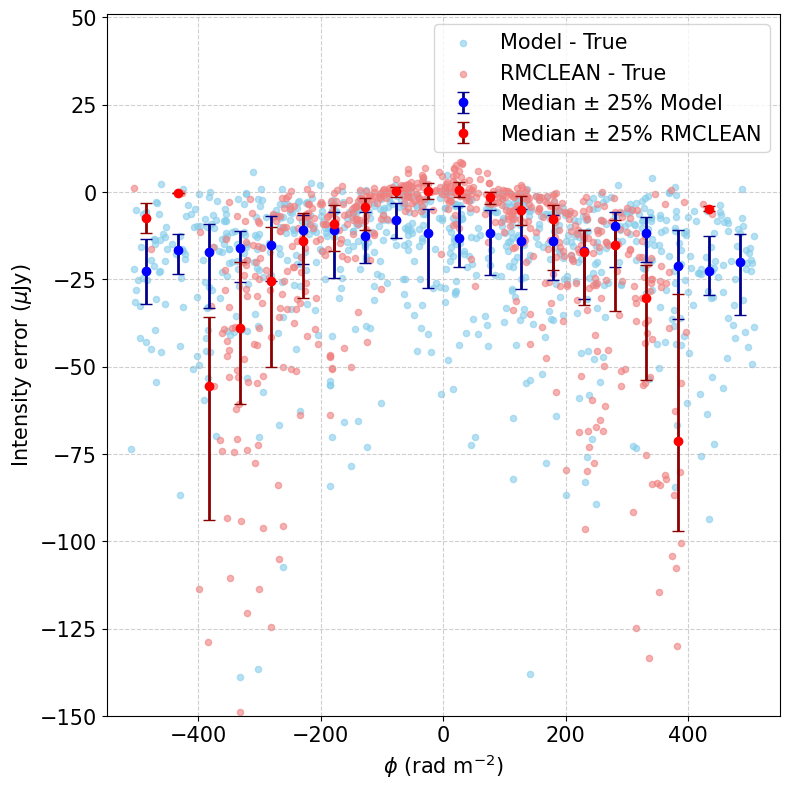}
    \caption{Sensitivity comparison of the deep learning model and RMCLEAN. The range is divided into 20 bins from which the median $\pm25\%$ are calculated and shown as dots and errorbars, respectively. While RMCLEAN shows better accuracy at low RMs, the deep learning model is able to find signals with higher RMs.}
    \label{fig:Depolarization}
\end{figure}

As mentioned is Sec.~\ref{sec:RM-synthesis}, the sensitivity of RM synthesis is dependent on the channel width and the RM of the source, as fastly rotating sources will lead to small net signals in broad channels. In Fig. \ref{fig:Depolarization} we compare the sensitivity of RMCLEAN and the deep learning model. The test is conducted with a single Gaussian with the rest of the parameters according to Table \ref{tab:sim_params2}, but with RMs up to $\pm512$ rad m$^{-2}$. While RMCLEAN demonstrates high accuracy at lower RMs, we observe that its sensitivity is significantly dependent on the RM value. After $\pm400$ rad m$^{-2}$, almost no signals are detected. The deep learning model has a lower accuracy at low RMs, but the sensitivity remains almost constant over the entire search range, reaching RMs up to 512 rad m$^{-2}$. As the model is given access to the intrinsic intensity of the source, it implicitly learns the sensitivity window of the observation, and adjusts the output accordingly. In total, the deep learning model and RMCLEAN are able to find 88\% and 72\%, respectively, out of the simulated signals.

\subsection{MeerKAT data of Abell 3376}

Next we proceede to test our model on the Abell 3376 dataset from the MGCLS.  RMCLEAN was run to a level of $3\sigma_{\text{P}}$ ($\sigma_{\text{P}}=4.3$ $\mu$Jy) with a gain of 0.10. To have a fair comparison between the models, both algorithms were run without masking any pixels. The deep learning model was run on a NVIDIA A100 GB GPU. The computation times for training and inference were 10 and 15 minutes, respectively, while RMCLEAN took 5 hrs. From the runtimes we can clearly see the benefits of parallel computing using GPUs, as this significantly reduces the overall computation time, enabling faster processing and analysis compared to traditional iterative CPU-based methods.  This is a key feature for larger sky surveys, where an automated RM deconvolution algorithm is necessary to handle the large data sizes efficiently.

\begin{figure*}
    \centering
    \includegraphics[width=\textwidth]{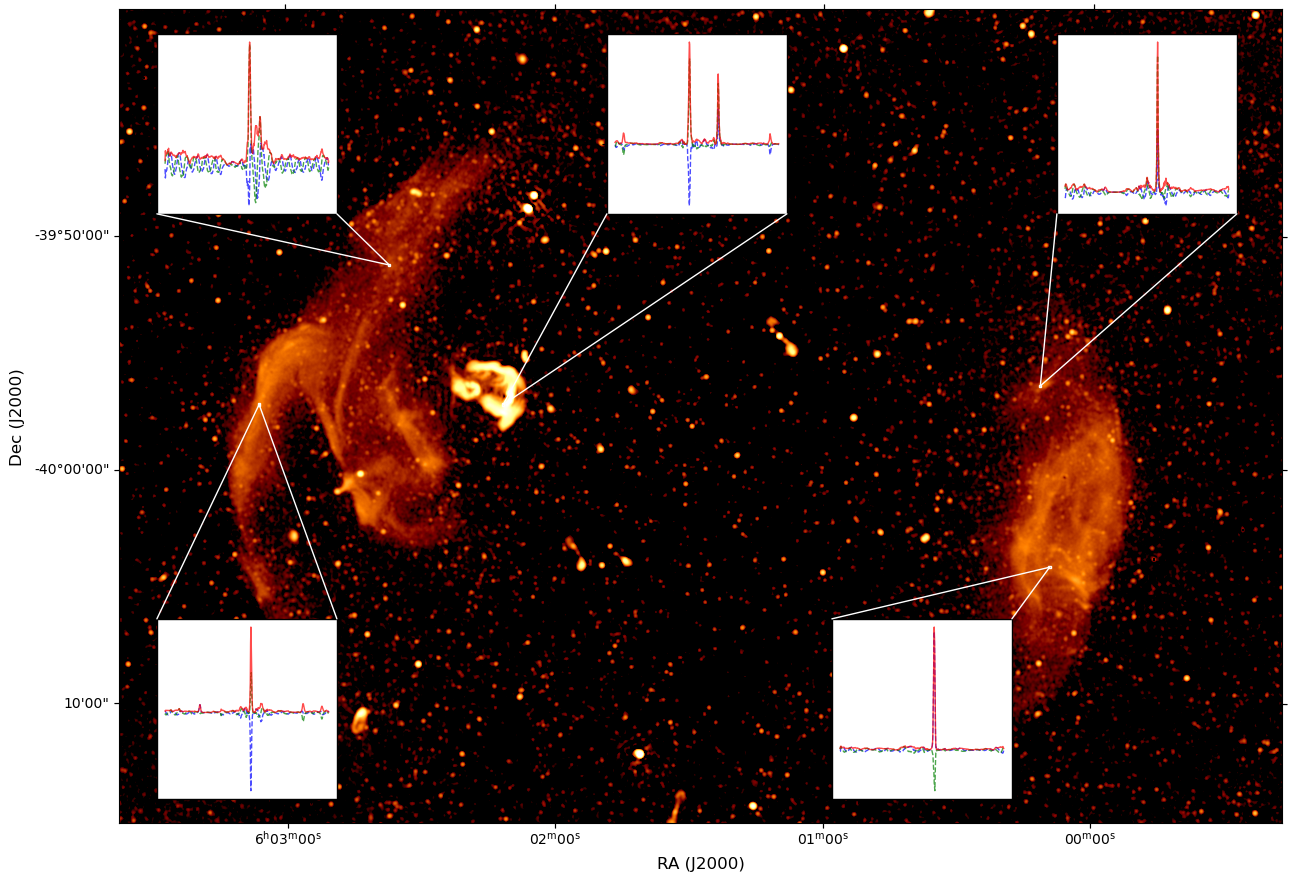}
    \caption{Total intensity map of Abell 3376 at a resolution of $8\arcsec\times8\arcsec$, with four example spectra from the cluster relics and one from the bright AGN. The spectra are color-coded as in Fig.~\ref{fig:RMCLEAN_1} and \ref{fig:RMCLEAN_2}, together with the magnitude shown in red.}
    \label{fig:spectrum_map}
\end{figure*}

In Fig.~\ref{fig:spectrum_map} we show how the Faraday dispersion varies in the relics and the bright AGN. In the plot five spectra are shown. Four come from line-of-sights through the radio relics in the cluster periphery, and one spectrum from the bright AGN close to the eastern relic. We observe that where there is strong emission, the RM synthesis signals are deconvolved to narrow peaks, indicating that the RM remains the same along the line of sight. In contrast, at points of weaker emission, typically closer to the cluster center, the Faraday spectrum is more complex. It is worth noting that the polarization cubes were contaminated by foreground emission, which dominates the regions beyond the cluster relics. Therefore, signals in these areas should be interpreted with caution. In the AGN, we observe a clear double peak signal, separated by about 100 rad m$^{-2}$.

\begin{figure*}
    \centering
    \includegraphics[width=\textwidth]{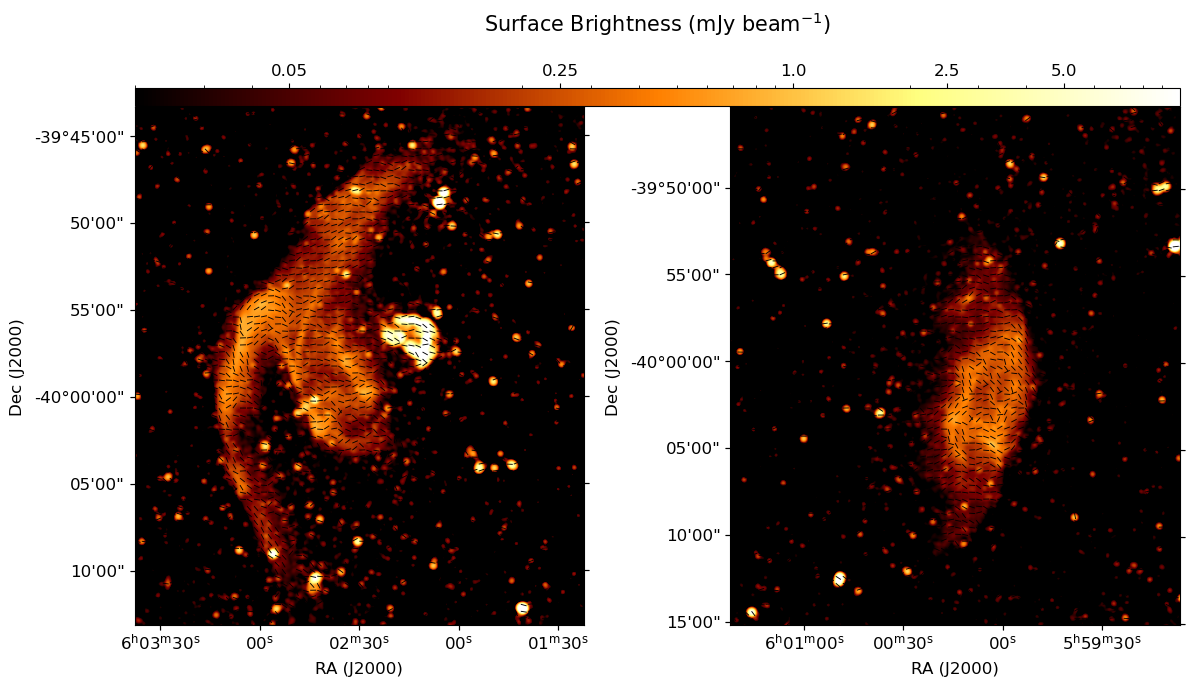}
    \caption{Total intensity map of the relics in Abell 3376 at a resolution of $15\arcsec\times15\arcsec$. Overlaid are the projected magnetic field vectors.}
    \label{fig:angle_map}
\end{figure*}

In Fig.~\ref{fig:angle_map} the orientation of the polarization vectors, rotated by 90 degrees to show the projected magnetic field orientation is shown. The image is produced after deconvolving the RM synthesis spectra according to the procedure in Sec.~\ref{sec:deep-learning} and extracting the intrinsic polarization angle of the polarized emission from the highest peak in $\phi$. We observe that the magnetic field lines along the outer edge of the cluster relics align with their shapes, particularly evident in the tail of the northeastern relic. This supports the theory that relics are a tracer of merging events, as the compression at the shock front is expected to align the magnetic field with it \citep{Ensslin_1998}. As we move inwards along the merger axis, the orientation becomes more turbulent and in the regions of weak emission, the orientation is seen as mostly random.

We observe that in most regions, the magnetic field orientation aligns with the dominant filament. However, there are some deviations from this trend, most notably at the upper part of the eastern relic tail (RA=$6^{\text{h}}03^{\text{m}}00^{\text{s}}$, Dec=$-40\degree00\arcmin00\arcsec$), where the magnetic field is oriented towards the inner part of the cluster. This is one of the discrepancies with \cite{Hu_2024}, where the orientation is perpendicular to the intensity gradient.

\begin{figure}[h]
    \centering
    \includegraphics[width=\columnwidth]{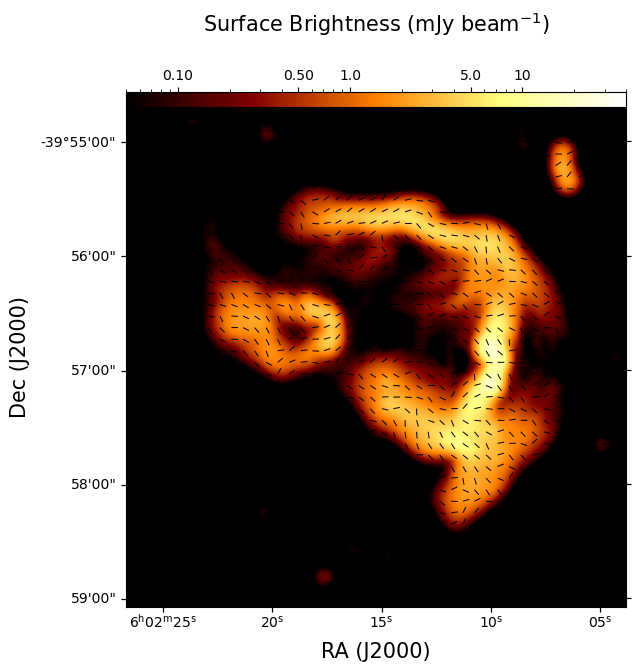}
    \caption{Total intensity map of the AGN west of the eastern relic, at a resolution of $8\arcsec\times8\arcsec$. Overlaid are the projected magnetic field vectors.}
    \label{fig:AGN}
\end{figure}

In Fig.~\ref{fig:AGN}, we present a close-up view of the radio galaxy MRC 0600-399 ($z=0.04559$). The image reveals that despite the jets bending from their initial trajectory, they stay collimated for approximately 100 kpc in the northern jet and 50 kpc in the southern jet beyond the bend point. Additionally, to the east of the galaxy, we observe a distinct structure, which, based on its higher redshift ($z=0.0480$), we identify as a separate galaxy \citep{Chibuezu_2021}. Furthermore, we see that the magnetic field orientation in both galaxies aligns with the jet outflow direction. 

\begin{figure}[h]
    \centering
    \includegraphics[width=\columnwidth]{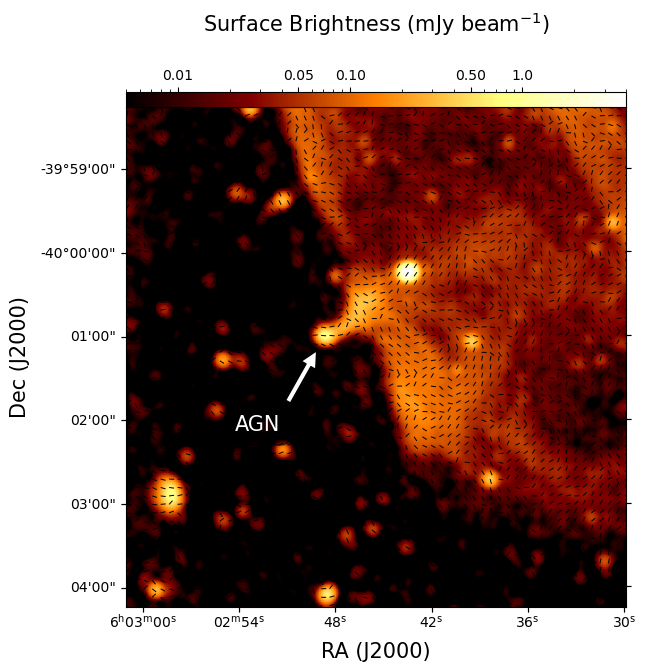}
    \caption{Total intensity map of the AGN connected to the eastern relic, at a resolution of $8\arcsec\times8\arcsec$. Overlaid are the projected magnetic field vectors.}
    \label{fig:AGN_2}
\end{figure}

In Fig.~\ref{fig:AGN_2} we show the AGN connected to the eastern relic. This AGN has been proposed by \cite{Chibueze_2023}, to provide the relic with seed electrons that potentially are re-accelerated by the shock. The link between the AGN and relic is confirmed by the magnetic field connecting the two structures. Additionally, there is a bright spot northwest of the protruding AGN. While the magnetic field in this region aligns with that of the relic, the Faraday spectrum reveals a complex source extended in $\phi$, in contrast to the narrow, well-defined peak in the surrounding region.

\begin{figure*}
    \centering
    \includegraphics[width=\textwidth]{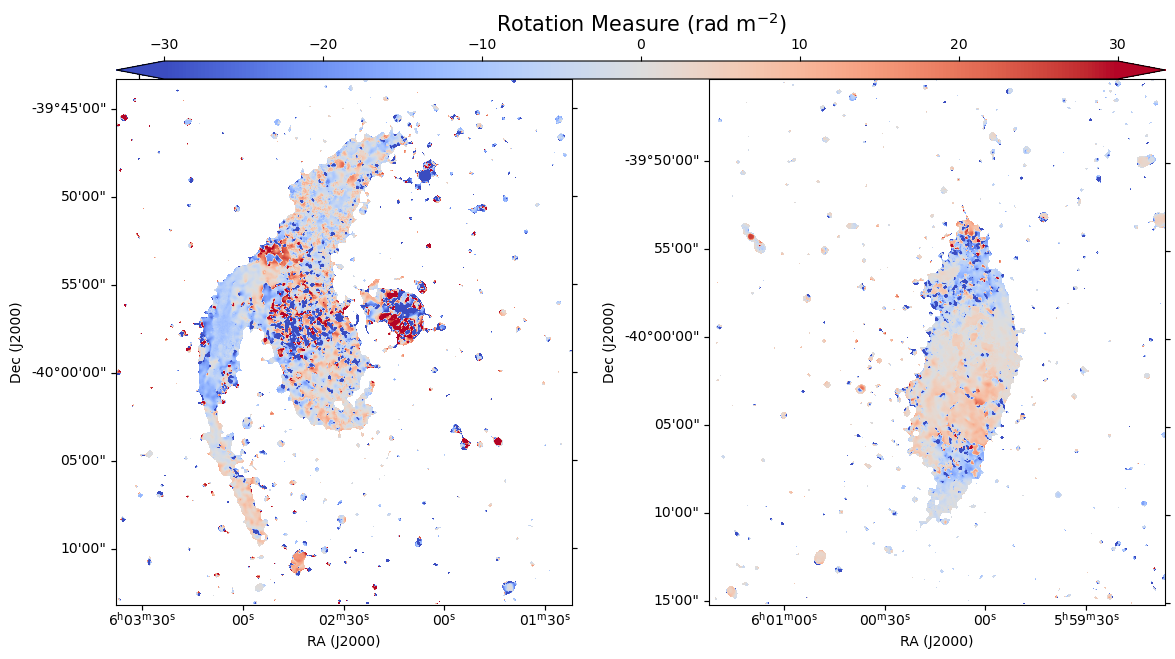}
    \caption{Rotation measure map of Abell 3376 at a resolution of $15\arcsec\times15\arcsec$. For contrast, the colorbar is clipped at $\pm30$ rad m$^{-2}$. Local estimates for the Galactic Faraday rotation have been subtracted, with $\phi = 22$ rad m$^{-2}$ for the eastern relic and $\phi = 32$ rad m$^{-2}$ for the western relic. Only pixels with a total intensity above $3 \sigma_{\text{I}}$ ($\sigma_{\text{I}}=13$ $\mu$Jy) are shown.}
    \label{fig:RM_map}
\end{figure*}

The rotation measure map of Abell 3376 is shown in Fig.~\ref{fig:RM_map}. The image is produced by localizing the highest peak in $\phi$ for each pixel, after which the galactic RM, taken from \cite{Hutschenreuter_2022}, is subtracted. For the most part of the relics, the RM appear to vary smoothly, except for the weak emission regions, where the peak emission RM values become very unstable, possibly indicating the sensitivity limit for the network. Furthermore, the RMs in the AGN show strong fluctuations, reaching values of up to $\pm50$ rad m$^{-2}$. 

\section{Discussion and Conclusions}
\label{sec:discussion}

In this section we discuss our results on, both, the simulated and observational data. We also discuss some implications of our deep learning model.

\subsection{Simulations}

In the simulations presented in Section \ref{sec:simulations}, we assumed a constant intrinsic polarization angle over each Gaussian component. Thus, for individual sources, the projected magnetic field would be aligned at each point of emission along the line of sight. Although this simplification might not fully capture the physical reality, modeling the intrinsic polarization variations along the line of sight is challenging. Furthermore, due to beam smearing, each pixel also contains information from neighboring pixels. In the case of a Faraday thick component with ordered magnetic fields, we would expect some kind of correlation of the polarization angle along $\phi$, while in the case of turbulent fields, the polarization angle can be completely random. It is crucial to note that the simulation outputs must still fit the observational data. If the initial polarization angle assumption is inaccurate, the model will adjust it accordingly. 

While RMCLEAN is able to locate the peak position of the spectra seen in Fig. \ref{fig:RMCLEAN_1} and \ref{fig:RMCLEAN_2}, it does not extract the intrinsic polarization angle in a correct manner for a scenario where the polarized intensity follows extended Gaussian profiles. As the algorithm extracts the real and imaginary parts of the Faraday dispersion function from the rapidly varying RM synthesis signal, the polarization angle is not expected to be correct outside of the peak location. This is seen in the secondary peaks in Fig.~\ref{fig:RMCLEAN_1} and \ref{fig:RMCLEAN_2}, where the real and imaginary parts do not correspond to the true signal. Furthermore, RMCLEAN uses the shift theorem of Fourier theory to rotate the RM synthesis signal to the average $\lambda^2$. While no information is lost in this transform \citep{Rudnick_2023}, it means that a small error in the peak position, will cause a significant error of the polarization angle when rotated back to $\lambda^2=0$ \citep{Brentjens_2005}.

From Fig.~\ref{fig:FWHM} we saw that the low resolution in Faraday depth, $\phi$, can be improved by deep learning. While there is no way of confirming the super-resolution on real data, where the true signal is unknown, we can enhance our confidence in the model's capabilities by using neighboring pixels as additional information. Since the model takes one-dimensional inputs and does not account for neighboring pixels, incorporating this spatial domain information should make the super-resolution results more robust.

In Fig.~\ref{fig:Depolarization}, we saw that depolarization effects, common in regular RM synthesis, can be reduced by learning the sensitivity window. However, caution should be taken when increasing the search range in $\phi$, as this can produce false positives in the output. This occurs because the first part of Eq. \ref{eq:loss_full} is RM sensitivity-dependent, as it includes the averaging operator.

\subsection{Abell 3376}
\label{sec:a3376}
In this section we discuss the results from applying the model to observational data, including some implications and suggestions for future work.

While Fig.~\ref{fig:angle_map} captures the magnetic field orientation on large scales, as the overlaid projected magnetic field vectors are an average produced from a number of pixels, the small-scale features reveal a different pattern. For example, in Fig.~\ref{fig:AGN_2}, we see that while the magnetic field is ordered on small scales, it is also tangled and complex, indicating turbulent processes at play. In this figure we can more clearly see how the scales on which the magnetic field is ordered are reduced as we move inwards into the relic. It thus becomes evident that the polarized fraction of the total flux should decrease due to beam depolarization effects. As the data is imaged using the CLEAN algorithm, the model image is convolved with a restoring beam to arrive at the final images. While the higher 8$\arcsec$ resolution cubes could have revealed finer structures of the magnetic field, the 15$\arcsec$ cubes were used in this work to improve the signal to noise at the regions of weak emission. Future telescopes like the Square Kilometer Array (SKA) will allow us to study the magnetic field of radio relics in even greater detail, together with recovering lower surface-brightness features of polarized emission, due to greater sensitivity together with a higher angular resolution, thus reducing the amount of beam depolarization.

From the peaks in Fig.~\ref{fig:spectrum_map}, we see that Gaussian peaks are favored by the model, most likely due to our Gaussian mixture model prior, rather than the true underlying signal. As physically realistic magnetic fields likely do not follow Gaussian profiles in $\phi$, future studies should investigate the model's ability to recover non-Gaussian features. From the upper left plot in Fig.~\ref{fig:spectrum_map}, we see that the flux increases towards higher $|\phi|$, which most likely is not the true reality of the relic magnetic field. These features might arise due to the model's confusion caused by the use of a restrictive prior. Including more realistic and non-Gaussian profiles in the simulated samples could thus make the model more robust and reduce the amount of unphysical features.

As discussed in Sec.~\ref{pre-processing}, we do not add noise to the mock training data. A consequence of not including noise in the simulated samples is that the model sometimes produces false positives. Since the model is only trained at producing sparse representations, the model puts all the noise in peaks at locations that agree with the data. Examples can be seen in the upper middle and lower left plot in Fig.~\ref{fig:spectrum_map}, where we have a low baseline, but small peaks separated from the main peaks. As the noise in the real and imaginary parts should ideally be Gaussian distributed, the noise of their magnitude should follow a Rician distribution, that is, positive definite and asymmetric when the Gaussian noise have zero mean. Usually, before doing any quantitative analysis on the signal flux, one should therefore subtract the Rician bias from the Faraday spectrum. In this work, however, the opposite effect was observed, where the model systematically underestimated the total flux of signals. As this effect was not observed when noise was removed from the synthetic data, we believe there could be two reasons behind this. Either the model overestimates the noise level, leading to a conservative bias in the flux estimation, or some of the signal flux is put into the noise peaks.

The decision of omitting noise was made for two main reasons: Firstly, accurately modeling the true noise statistics is difficult. Our model assumes that the noise is Gaussian with zero mean and uncorrelated in $\lambda^2$. However, in radioastronomical images the noise is often correlated and not Gaussian. Thus, adding purely Gaussian noise led to a decreased model performance, as the model struggled to transfer knowledge from simulations to real data and vice versa. We also attempted to create more realistic noise by extracting samples without significant signal from the data and then inserting a simulated signal into those samples. While this approach helped to reduce false positives in the outputs, it falls short of accurately reproducing the data.

The second reason concerns the loss function, which becomes problematic when incorporating noise in the simulations. The first component of Eq. $\ref{eq:loss_full}$ seeks to minimize the difference between the output in $\lambda^2$ and the observed data, including the noise. However, the second component aims to replicate the underlying signal, excluding the noise. Thus, the two loss terms compete against each other, and requires careful tuning of the $\beta$ parameter in order for the model to benefit from both terms. By excluding noise in the simulated samples, both components of the loss function should aim to find a solution to the data, regardless of the noise statistics.

Furthermore, it is possible that Eq.~\ref{eq:measurement_eqation} does not accurately capture the true measurement operation. One could thus draw inspiration from various techniques used in interferometric imaging, where terms are included to account for variations in the primary beam and the $w$-component. Similar effects in RM synthesis would include varying channel widths and bandwidth depolarization effects.

While this work has focused on polarization data from the MGCLS, the model should perform similarly well in other frequency bands and with other telescopes. One only has to change the RM search range and resolution, as well as the FWHM of the sources that one expects to find. While it is possible to train the network on data from a range of instruments, we expect higher performance when tailored to a specific spectral band. Furthermore, the option of training on one dataset and inferring on another has not been studied in this work. This should in theory not be an issue, as long as the two datasets have a similar dynamic range and noise level. One could also create a training dataset from a variety of datasets which collectively represent a wide range of conditions, thereby improving the robustness and generalizability of the model.

\subsection{Summary}
\label{sec:summary}

In this paper, we have developed a deep learning model to perform RM synthesis deconvolution. The model was trained using a Gaussian mixture model prior alongside observational data from the MeerKAT Galaxy Cluster Legacy Survey DR1. It is important to note that only 20\% of the training data was simulated, meaning that while the simulated dataset guides the model to some extent, the majority of information is derived from the observational dataset, ensuring the model remains primarily influenced by real-world data. We tested the model on simulations as well as observational data from the MeerKAT galaxy cluster data. The model was able to recover complex and high-RM signals better than the traditional CLEAN algorithm, in a less computationally expensive way. For future studies, it would be beneficial to compare the deep learning model with an algorithm designed to deconvolve Faraday thick components, such as a multi-scale variant of RMCLEAN or a multi-component adaptation of $QU$-fitting. When applied to data from the Abell 3376 galaxy cluster, the model was able to deconvolve Faraday spectra from various parts of the intra-cluster medium, and map the magnetic field structure of the cluster in fine detail. We found that while the dominant filament largely determines the magnetic field orientation, there are regions where this assumption does not hold. One should thus be cautious when using methods like the synchrotron intensity gradient mapping, as studying clusters in full polarization can yield more comprehensive information. The model was also applied to five other datasets from the MGCLS, out of which only two has previously been studied with MeerKAT in full Stokes. The results generally agrees with those from previous studies, with the main difference being the great sensitivity of the MeerKAT telescope, providing a great amount of detail to e.g. the Abell 85 cluster where we have mapped the magnetic field structure of a complex phoenix. Furthermore, the model was able recover the magnetic field orientation from regions of weak emission from e.g. the Abell 3667 cluster, revealing ordered magnetic fields on Megaparsec scales. The high accuracy and low computational cost of the model makes it a good candidate for forthcoming surveys, where an automated pipeline will be crucial to handle the large data sizes.

\begin{acknowledgements}

VG acknowledges support by the German Federal Ministry of Education and Research (BMBF) under grant D-MeerKAT III.
MB acknowledges funding by the Deutsche Forschungsgemeinschaft (DFG, German Research Foundation) under Germany's Excellence Strategy -- EXC 2121 ``Quantum Universe'' --  390833306.
The MeerKAT telescope is operated by the South African Radio Astronomy Observatory, which is a facility of the National Research Foundation, an agency of the Department of Science and Innovation. The authors acknowledge the contribution of those who designed and built the MeerKAT instrument.

This work made use of Astropy:\footnote{http://www.astropy.org} a community-developed core Python package and an ecosystem of tools and resources for astronomy \citep{2022ApJ...935..167A}.

\end{acknowledgements}

%
%

\bibliography{ref}
\bibliographystyle{aa}

\begin{appendix}

\section{Results from other galaxy clusters observed with MeerKAT}
\label{sec:clusters}

We also applied our deep learning deconvolution model to other clusters from the MGCLS DR1. For some of these clusters RM data has already been published and for others we show the polarization data for the first time. In this appendix we present results from some interesting sources. A more detailed physical interpretation is going to be presented in separate papers.

\begin{figure*} 
    \centering
    \includegraphics[width=\textwidth]{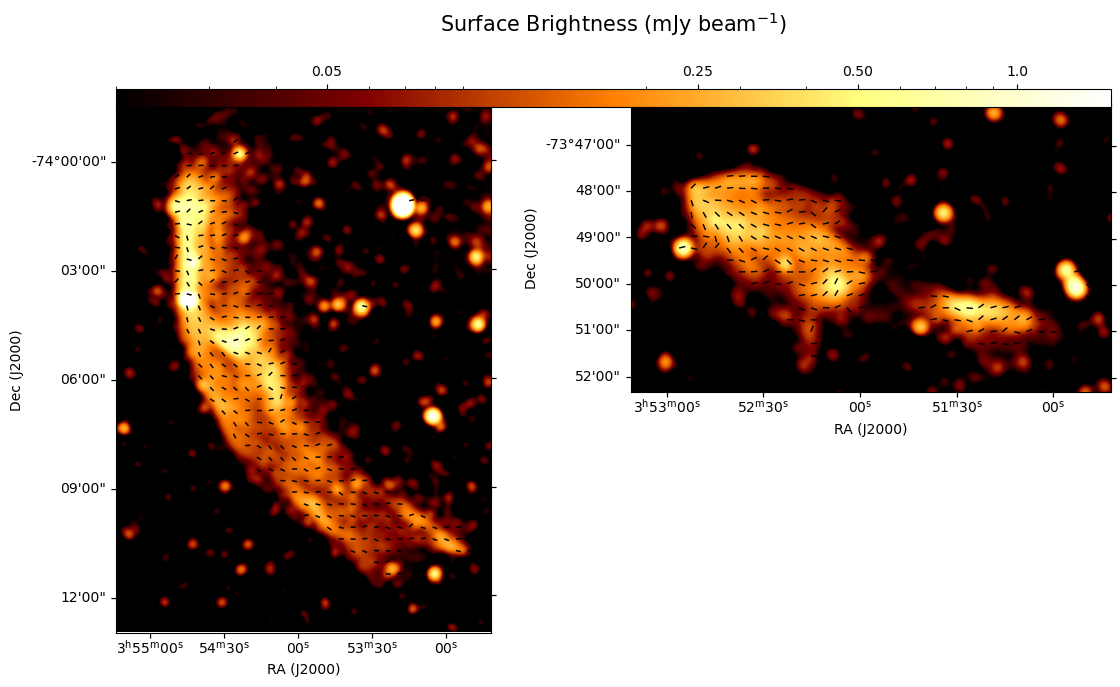}
    \caption{Radio emission from the Abell 3186 galaxy cluster relics at a resolution of $15\arcsec\times15\arcsec$, overlaid with the polarization $B$-vectors.}
    \label{fig:A3186}
\end{figure*}

\paragraph{Abell 3186}
The cluster Abell 3186 (also called MCXC J0352.4-7401) has been studied by e.g. \cite{Duchesne_2021} using the Murchison Widefield Array (MWA) and ASKAP telescopes. Most recently it has been observed by \cite{Hu_2024} with MeerKAT, using the SIG method to study the magnetic field structure of the relics. In Fig.~\ref{fig:A3186} the two radio relics in Abell 3186 are shown. We see that in both relics the magnetic field aligns well with the orientation of the relics, particularly near the shock front. Overall, the results align well with \cite{Hu_2024}, as the polarization vectors generally follow the orientation of the dominant filament. However, at the inner edge of the south-western relic, some differences are noticeable. In Fig.~\ref{fig:A3186}, the magnetic field is oriented towards the inner region of the cluster, while the SIG method consistently results in field vectors perpendicular to the local gradient. A similar pattern is observed in the north-eastern relic, although it is less pronounced. These findings suggest that SIG might not capture certain details revealed by polarization studies, particularly when the magnetic field exhibits unexpected orientations. Furthermore, studying the polarization properties at a pixel-by-pixel basis allow us to obtain a unbiased result, independent on the surrounding region.

\begin{figure*} 
    \centering
    \includegraphics[width=\textwidth]{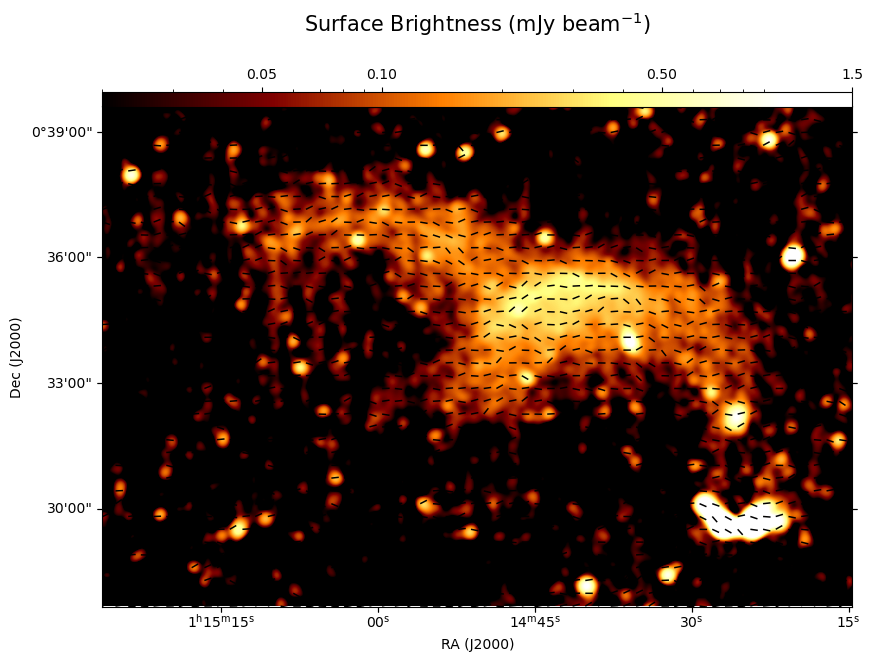}
    \caption{Radio emission from the Abell 168 galaxy cluster relic at a resolution of $15\arcsec\times15\arcsec$, overlaid with the polarization $B$-vectors.}
    \label{fig:A168}
\end{figure*}

\paragraph{Abell 168}
Next we show the results from the cluster Abell 168. This double relic cluster has previously been studied by e.g. \cite{Dwarakanath_2018} using the Karl G. Jansky Very Large Array (VLA) and the Giant Meterwave Radio Telescope (GMRT). However, polarization studies have not been conducted of the cluster. In Fig.~\ref{fig:A168} the cluster is shown together with the projected magnetic field orientation. While an arc-shaped relic is located, the second relic found by \cite{Dwarakanath_2018} is not visible. The reason for this is that regions far from the pointing center are blanked in the MGCLS enhanced products, which is the case for the relics in this cluster. In the relic detected here it is striking how uniform the magnetic field orientation is throughout the relic. At least at our angular resolution it does not show the variance that is observed in some of the other relics.

\begin{figure*} 
    \centering
    \includegraphics[width=\textwidth]{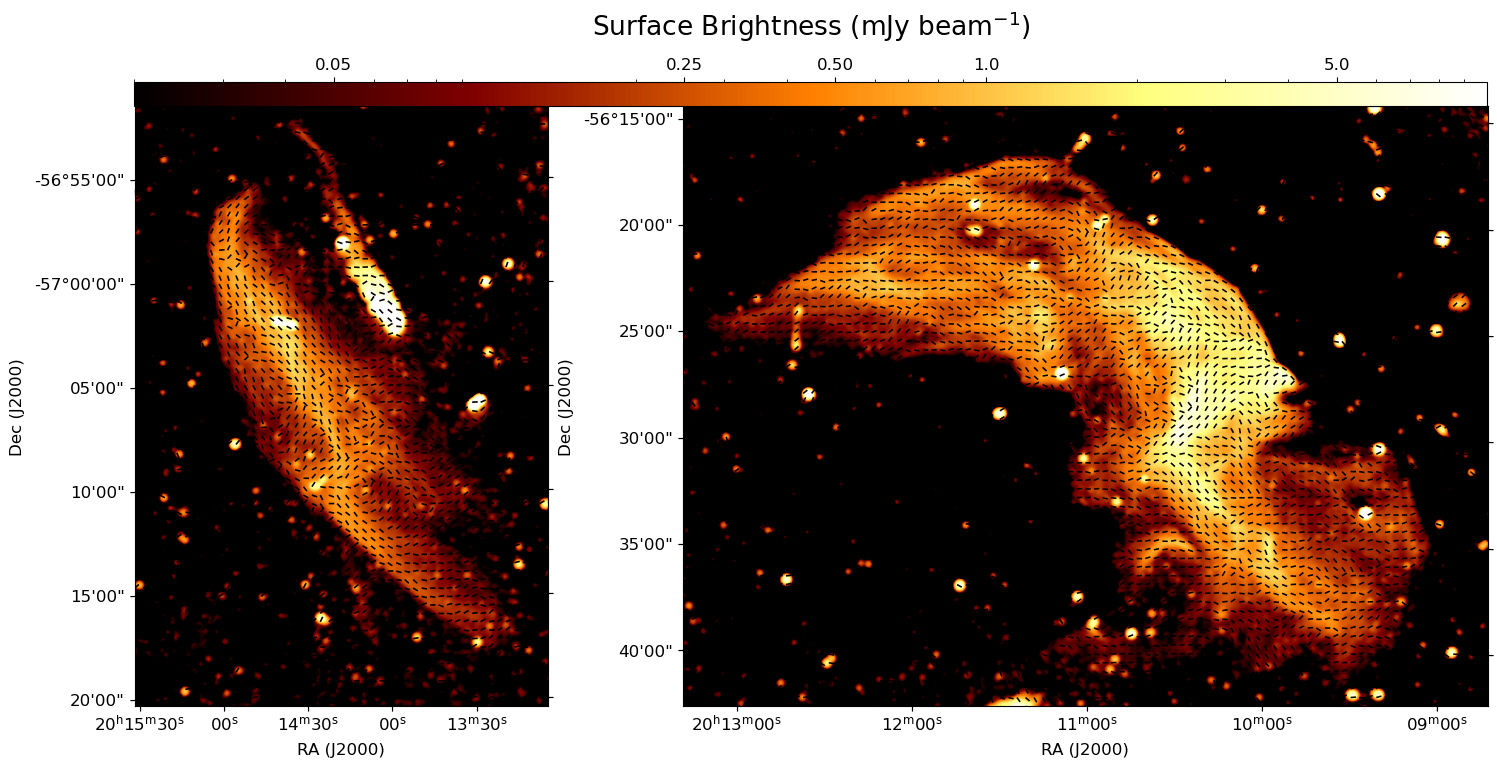}
    \caption{Radio emission from the Abell 3667 galaxy cluster relics at a resolution of $15\arcsec\times15\arcsec$, overlaid with the polarization $B$-vectors.}
    \label{fig:A3667}
\end{figure*}

\paragraph{Abell 3667}
The cluster Abell 3667  has recently been studied extensively in full polarization with MeerKAT L-band by \cite{de_Gasperin_2022}. The authors discovered that the relics are composed of a network of synchrotron filaments with varying spectral and polarization properties. These are likely linked to multiple regions of particle acceleration and localized magnetic field enhancements. In Fig. \ref{fig:A3667} we show the magnetic field orientation of the two radio relics. We see that the magnetic field is ordered on scales comparable to the relic extension, and that the main contributor to the field orientation is the local orientation of the dominant filament. When comparing our results with \cite{de_Gasperin_2022}, we find that the magnetic field in our analysis appears to remain ordered even in regions of weak emission, whereas \cite{de_Gasperin_2022} indicates turbulent magnetic fields in these areas. Since both studies are based on the same set of visibilities, the differences might stem from either the CLEANing process to a higher resolution or the thresholding methods used by \cite{de_Gasperin_2022}. Furthermore, the authors use a Faraday depth sampling of 2 rad m$^{-2}$, as opposed to 1 rad m$^{-2}$ in this work. As the regions of weak emission are expected to be Faraday depolarized, a small change in $\phi$ would potentially result in a great difference in the observed polarization angle. Additionally, the authors used a Högbom algorithm similar to RMCLEAN to produce the polarization maps, but it is uncertain if this is the cause of the divergent results.

\paragraph{Abell 85}

\begin{figure*} 
    \centering
    \includegraphics[width=\textwidth]{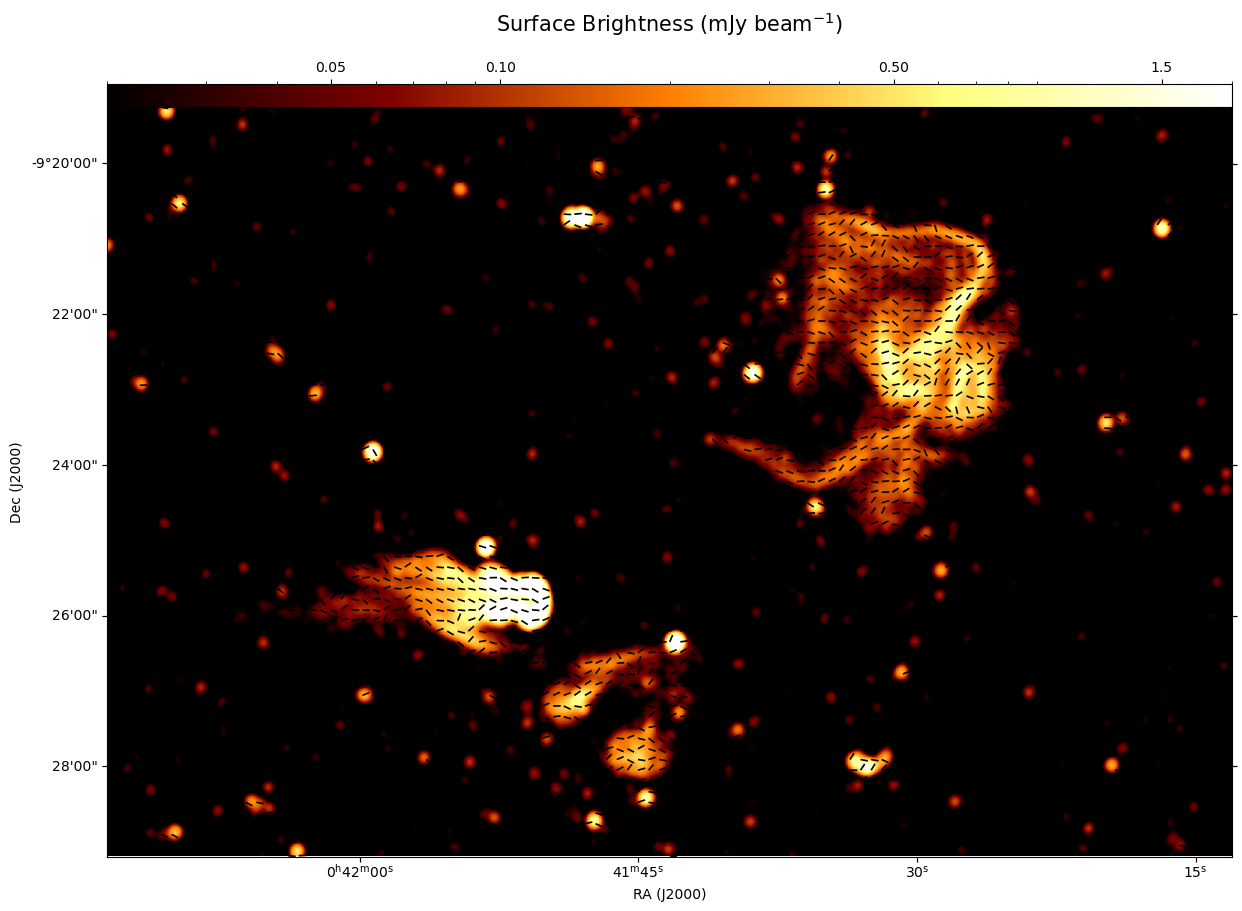}
    \caption{Radio emission from the galaxy cluster Abell 85 at a resolution of $8\arcsec\times8\arcsec$, overlaid with the polarization $B$-vectors.}
    \label{fig:A85}
\end{figure*}

\begin{figure}[ht]
    \centering
    \includegraphics[width=\columnwidth]{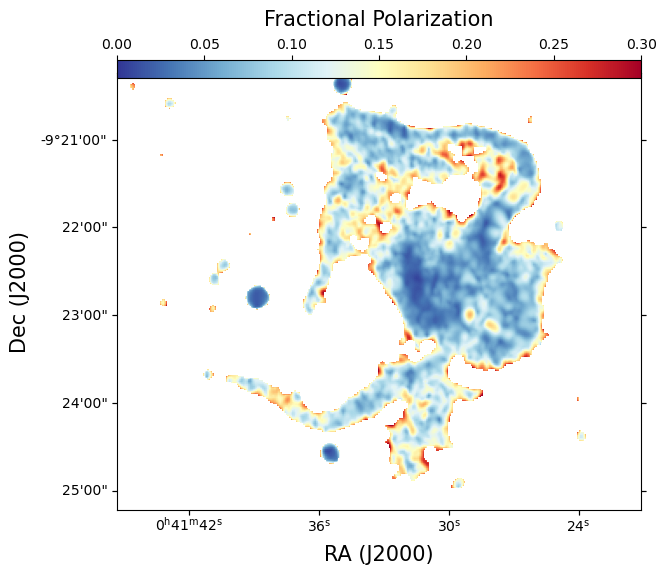}
    \caption{Fractional polarization map (908-1656 MHz) of the phoenix in Abell 85, at a resolution of $8\arcsec\times8\arcsec$. Only pixels with a total intensity above 5$\sigma_I$ ($\sigma_I$ = 12 $\mu$Jy) are shown.}
    \label{fig:frac_pol_A85}
\end{figure}

In Fig.~\ref{fig:A85} we show a cutout from the cluster Abell 85, displaying some interesting sources. Note that we did not include the radio halo at the center of the cluster that was found by \cite{Knowles_2022} because its emission is very weak. Most notably in our map to the north-west we find a complex phoenix (revived fossil plasma source), which has previously been studied by, e.g., \cite{Slee_1984}, \cite{Giovannini_2000} \cite{Duchesne_2021_2}. However, the classification varies between a phoenix and a relic. It is interesting to note that in the northern and southern filaments the magnetic field vectors follow the geometry of the filaments. Furthermore, south-east of the phoenix, we identify an AGN with wide angle jets, whose magnetic field orientation align well with the outflow direction. \cite{Slee_2001} could not determine the field vectors because they could not Faraday-derotate their data. Also they only saw the central and northern parts of this phoenix. But they found significant variation in the polarization fraction in A85, ranging from 35\% in the northern part to 10\% in the central part of the phoenix (they label it as south-eastern arc). In Fig.~\ref{fig:frac_pol_A85} the fractional polarization map from the MGCLS is shown. The map is produced from the mean of $\sqrt{Q^2 + U^2}$ over the entire band. As the resulting polarized flux is not corrected for Rician bias, the polarized fractions should be considered as upper limits. Similar to \cite{Slee_2001} we find a polarization fraction of about 10\% in the central part of the phoenix, but with some regions below 5\%. In the filaments the polarized fraction is generally higher, around 10-15\%, with some hotspots with a fractional polarization between 20\% and 30\%.

\begin{figure}[ht]
    \centering
    \includegraphics[width=\columnwidth]{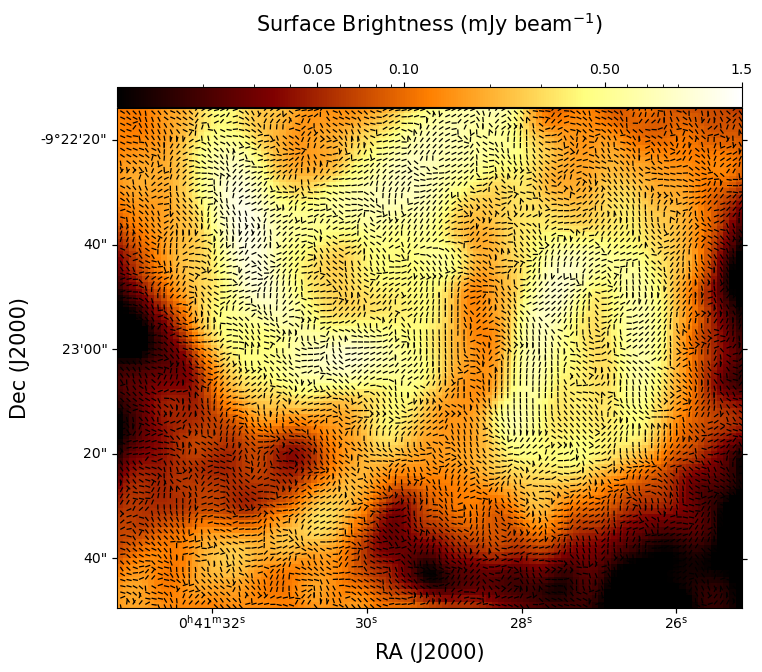}
    \caption{Radio emission from the central region of the phoenix in Abell 85 at a resolution of $8\arcsec\times8\arcsec$, overlaid with the polarization $B$-vectors.}
    \label{fig:angle_map_A85_inner}
\end{figure}

In Fig.~\ref{fig:angle_map_A85_inner} the magnetic field orientation of the inner region of the phoenix is shown, in order to highlight the vortex-like structures. We see that the magnetic field of this region is very complex and in some regions ordered on just the scale of the angular resolution. In the regions of the bright torus structures, the field lines appear to follow the orientation of the filament, while where there is weak emission the magnetic field is more chaotic. These rapid rotations of the magnetic field are likely the cause of the depolarization effects seen in the central region of Fig.~\ref{fig:frac_pol_A85}. \cite{Ensslin_2002} proposed that the torus structures seen in Fig.~\ref{fig:angle_map_A85_inner} are formed by light radio plasma moving through a shock wave. At the shock front, the ram pressure of the pre-shock gas and the thermal pressure of the post-shock gas are in balance. When the lighter radio plasma comes in contact with the shock front, the ram pressure is reduced at the point of contact. This causes the post-shock gas to expand into the volume occupied by the low-pressure radio plasma. This process disrupts the radio plasma, eventually forming a torus structure, similar to a smoke-ring. The similarity between the simulations of \cite{Ensslin_2002}, and the magnetic field in Fig.~\ref{fig:angle_map_A85_inner} strengthen the argument that these toroidal structures are a result of such a compression scenario.

\paragraph{Abell 194}

\begin{figure*} 
    \centering
    \includegraphics[width=\textwidth]{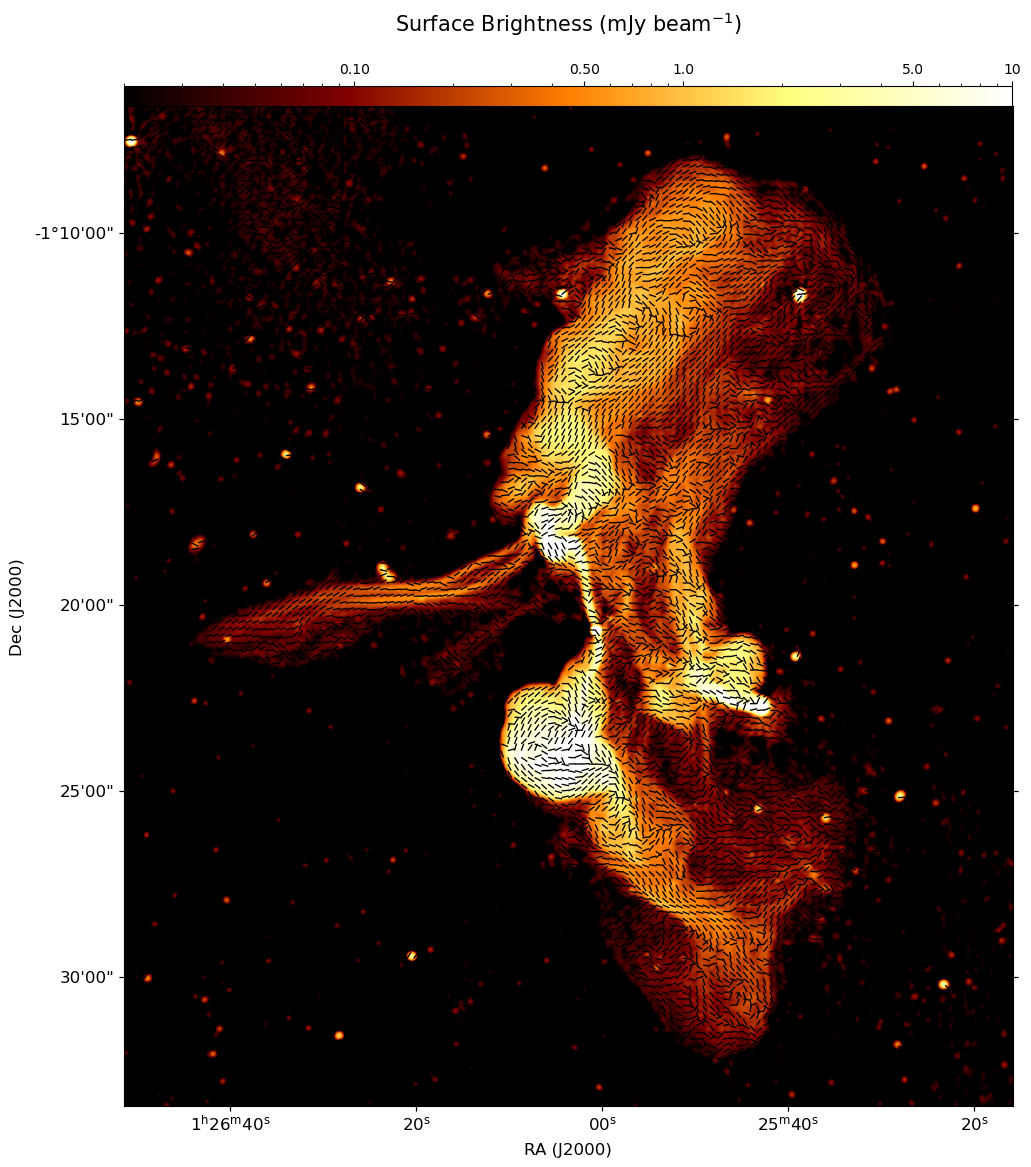}
    \caption{Radio emission from two radio galaxies in the galaxy cluster Abell 194 at a resolution of $8\arcsec\times8\arcsec$, overlaid with the polarization $B$-vectors.}
    \label{fig:A194}
\end{figure*}

\begin{figure*}[!htb]
    \centering
    \begin{subfigure}{\textwidth}
        \centering
        \includegraphics[width=\textwidth]{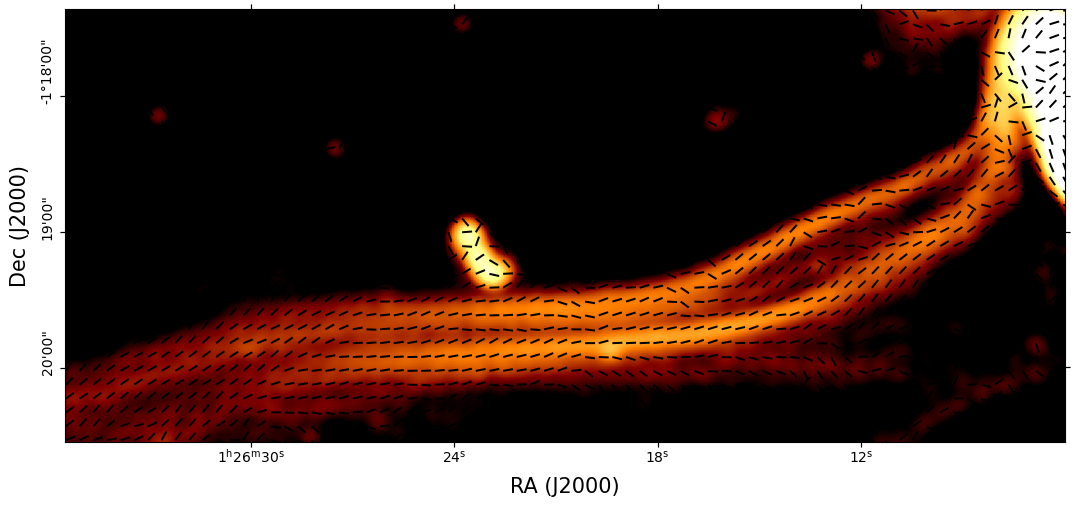}
    \end{subfigure}
    
    \begin{subfigure}{\textwidth}
        \centering
        \includegraphics[width=\textwidth]{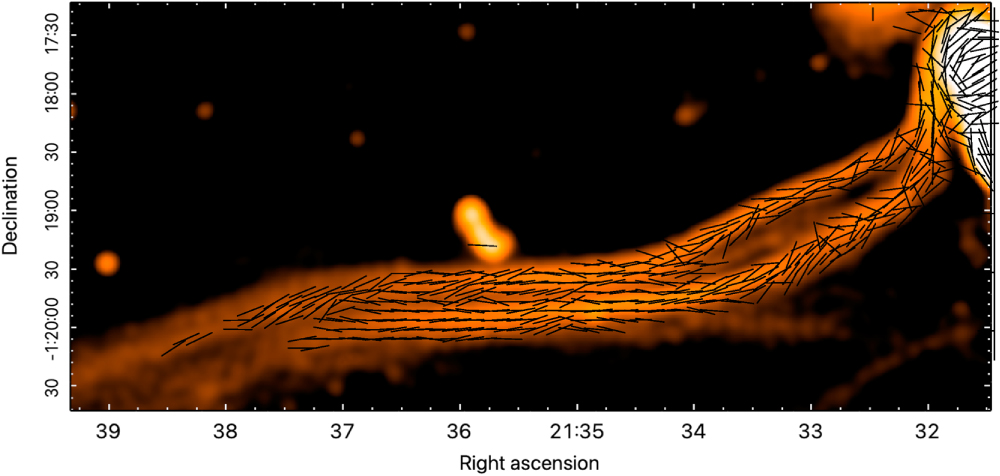}
    \end{subfigure}
    
    \caption{Comparison between this work (top) and \cite{Rudnick_2022} (bottom), of the magnetic field orientation in the long eastern filament.}
    \label{fig:Filament}
\end{figure*}

Finally, in Fig.~\ref{fig:A194} we show the radio emission together with the polarization $B$-vectors for two radio galaxies in the galaxy cluster Abell 194. The same dataset has previously been studied by \cite{Rudnick_2022}, including an extensive RM study of the two radio galaxies, together with a map of the magnetic field of the long eastern filament. The eastern filaments are  around 50\% polarized, with no detectable net RM with respect to the Galactic foreground, and only small rms variations (9 rad m$^{-2}$) along their length. This enables a mapping between Faraday depth and distance along the line-of- sight. The magnetic field vector is shown to follow the length of the filament, supporting the interpretation that these are magnetic flux tubes. In Fig.~\ref{fig:Filament} a comparison between this work and \cite{Rudnick_2022} is shown for the filament. We see that our results are similar to those of \cite{Rudnick_2022}. The field vectors in our work seem to vary more smoothly over the filament. However, this might just be to different vector lengths and number of samples used to produce such vectors.

\end{appendix}

\end{document}